\documentclass[aps,10pt,a4paper,showpacs,twocolumn,pre]{revtex4-1}

\usepackage{amssymb}
\usepackage{mathrsfs}
\usepackage{stmaryrd}
\usepackage[dvips]{graphicx}
\usepackage{subfigure}
\usepackage[dvips]{color}
\usepackage{caption}
\newcommand{\mb}{\mathbf}
\newcommand{\mc}{\mathcal}

\begin{document}

\title{Stimuli-responsive brushes with active minority components: Monte Carlo study and analytical theory}

\author{Shuanhu Qi}
\email{qish@uni-mainz.de}
\affiliation{Institut f\"{u}r Physik, Johannes Gutenberg-Universit\"{a}t Mainz, Staudingerweg 7, D-55099 Mainz, Germany}

\author{Leonid I. Klushin}
\email{leo@aub.edu.lb‎}
\affiliation{Department of Physics, American University of Beirut, P. O. Box 11-0236, Beirut 1107 2020, Lebanon}

\author{Alexander M. Skvortsov}
\email{astarling@yandex.ru}
\affiliation{Chemical-Pharmaceutical Academy, Professora Popova 14, 197022 St. Petersburg, Russia}

\author{Alexey A. Polotsky}
\email{alexey.polotsky@gmail.com}
\affiliation{Institute of Macromolecular Compounds of the Russian Academy of Sciences, 31 Bolshoy pr., 199004 St. Petersburg, Russia}

\author{Friederike Schmid}
\email{friederike.schmid@uni-mainz.de}
\affiliation{ Institut f\"{u}r Physik, Johannes Gutenberg-Universit\"{a}t Mainz, Staudingerweg 7, D-55099 Mainz, Germany}

\bigskip

\begin{abstract}

Using a combination of analytical theory, Monte Carlo simulations, and three
dimensional self-consistent field calculations, we study the equilibrium
properties and the switching behavior of adsorption-active polymer chains
included in a homopolymer brush. The switching transition is driven by a
conformational change of a small fraction of minority chains, which are
attracted by the substrate. Depending on the strength of the attractive
interaction, the minority chains assume one of two states: An exposed state
characterized by a stem-crown-like conformation, and an adsorbed state
characterized by a flat two-dimensional structure. Comparing the Monte Carlo
simulations, which use an Edwards-type Hamiltonian with density dependent
interactions, with the predictions from self-consistent-field theory based on
the same Hamiltonian, we find that thermal density fluctuations affect the
system in two different ways. First, they renormalize the excluded volume
interaction parameter $v_\mathrm{\tiny bare}$ inside the brush. The properties of the brushes can
be reproduced by self-consistent field theory if one replaces $v_\mathrm{\tiny bare}$ by an
effective parameter $v_{\mbox{\tiny eff}}$, where the ratio 
of second virial coefficients $B_{\mbox{\tiny eff}}/B_\mathrm{\tiny bare}$
depends on the range of monomer interactions, but not on the grafting density,
the chain length, and $v_\mathrm{\tiny bare}$.  Second, density fluctuations affect
the conformations of chains at the brush surface and have a favorable effect on
the characteristics of the switching transition: In the interesting regime
where the transition is sharp, they reduce the free energy barrier between the
two states significantly. The scaling behavior of various quantities is also
analyzed and compared with analytical predictions. 

\end{abstract}

\maketitle

\section{Introduction}

Polymer brushes are organic surface layers formed by polymer chains that are
grafted at one end to a substrate \cite{Lodge,Milner}.  Since the thickness of
brush layers is typically in the nanometer range, they are interesting for the
design of functional surfaces \cite{Russell,R1,R2} with applications in a wide
variety of areas ranging from colloidal stabilization
\cite{colloidal_stabilization}, lubrication \cite{lubrication}, controlled
friction and adhesion \cite{adhesion}, anti-fouling \cite{antifouling},
biocompatibility\cite{biocompatibility}, drug delivery \cite{drug_delivery},
and smart stimuli-responsive materials \cite{R1}. In this respect,
multicomponent polymer brushes are particularly promising \cite{R1,R2}. If the
brush chains are covalently bound to the substrate, they cannot phase separate
on a global scale, but they can still develop structure on the nanoscale.  The
resulting brush morphologies are controlled both by the intrinsic properties of
the brush, e.g., chemical properties of the chains (compatibility or
incompatibility), the grafting densities, the chain lengths, and by the
environment-related parameters, such as solvent selectivity, substrate
preference, temperature, and the pH. Due to the ability of polymer brushes to
selectively respond to environmental stimuli, they can be used to design
materials that can reversibly switch/tune their surface properties, e.g., with
respect to wettability \cite{wettability,wettability_time}, permeability
\cite{permeability}, friction \cite{friction}, and optical properties
\cite{optical}.

Stimuli-induced phase separation provides the basic mechanism
for the change in the brush surface properties depending on which
of the two microphases forms the outer part of the brush. A typical
example for such a morphology-related switchable surface is a mixed
polymer brush with equal amounts of hydrophobic and hydrophilic polymers
grafted on a substrate. By treatment of different solvents, the surface
composition can change and then the wettability of this material switches
\cite{wettability,wettability_time}. Morphology change necessarily
involves slow highly cooperative chain dynamics and therefore a typical
response time turns out to be in the range of several minutes or larger
\cite{wettability_time}.

In a recent letter, we have proposed a new class of brush-based
switches \cite{Klushin_switch}, which rely on a radical conformational
change of individual adsorption-active minority chains in an otherwise
inert brush.

The selective adsorption driving the transition may arise,
e.g., from electrostatic interactions or hydrogen bonding between
active groups in the minority chains and the substrate. These active
groups can serve as responsive sensors that trigger the switching
transition. The most radical change is associated with the end group
of the minority chain. In the adsorbed state it resides in close proximity
of the substrate deeply buried within the brush, while in the other
state it is exposed to the environment. Thus the transition could
potentially promote a specific immune-like response. Chemically
or biologically active groups can be attached to the free end of the
minority chain to serve as practically useful sensors.

Based on theoretical arguments and simple one-dimensional mean-field
calculations, we  demonstrated that even a small chain length increment of
about 10 \% produces a sharp transition from the exposed state to the adsorbed
state. The transition time is expected to be very short since the free energy
barrier between these two thermodynamically stable states is about several
$k_{B}T$.  A further increase in the minority chain length will lead to sharper
transitions, but at the same time to longer transition times due to higher
barriers. The strong response of the chain conformation of a minority chain
\cite{brush_minority_1997} to small variations of its length is related to
the recently reported ``surface instabilities'' in polymer brushes
\cite{Merlitz1,Romeis1}, which can be used to sense solvent quality 
\cite{Merlitz2,Romeis2}. In our work, we proposed to use single
chains as switches triggered by a change in substrate-polymer interaction. One
major benefit of this switch is that it does not involve cooperative
rearrangements of many chains, therefore the switching transition is fast as
compared to the existing examples in mixed brushes. 

The arguments presented in Ref.\ \cite{Klushin_switch} rely
on several assumptions. First, the presence of the minority chains
was taken to have no effect on the surrounding polymer brush. Second,
thermal fluctuations of the majority brush component were disregarded.

In the current paper, we present extensive Monte Carlo (MC)
simulations of a coarse-grained model for polymer brushes with a single
immersed adsorption-active minority chain. To assess the influence
of fluctuations separately, we have also performed three-dimensional
self-consistent field (SCF) calculations for comparison. We investigate
in detail both lateral and longitudinal
characteristics of chain conformations and the various characteristics
of the transition.

The remainder of the paper is organized as follows: Sec.\ \ref{sec:MC_model}
outlines the MC model and the simulation scheme. Sec.\ \ref{sec:brush} presents
the MC results for a homogeneous monodisperse brush and comparison with the
established analytical SCF theory, and analyzes the dependence of the
renormalized virial coefficient on the effective interaction range. Sec.\
\ref{sec:theory} presents a more detailed derivation of the theory for the
adsorption-active minority chain sketched in Ref.\ \cite{Klushin_switch}.
In Sec.\ \ref{sec:results} the main results of the MC simulations are presented
and discussed. This section also compares MC and 3-d SCF calculations with the
emphasis on fluctuation effects. A general discussion is given in the final
section\ \ref{sec:conclusions}.

\section{MC simulation model}\label{sec:MC_model}

In the simulation community, a variety of
numerical methods have been developed and used to investigate polymer
brush systems, ranging from molecular dynamics \cite{MD1}, dissipative
particle dynamics \cite{DPD1}, and Monte Carlo (MC) simulations \cite{MC1,MC_CG,MC3,MC4,MC5}
to numerical self-consistent field (SCF) calculations \cite{lattice_SCF,SCF2,SCF3,SCF4,SCF5,SCF6}.
The present study mainly relies on MC simulations, while for comparison
we also present results obtained from 3-d SCF theory.
The MC simulation model and scheme are described as follows, and the
detailed description of SCF method is shown in the Appendix.

In the MC simulations, we adopt a coarse-grained off-lattice model first
proposed by Laradji et al \cite{MC_CG}. In this approach, a particle-based
representation of the polymers is combined with an Edwards type Hamiltonian
\cite{EdwardsType1,EdwardsType2}, which defines non-bonded interactions in
terms of local monomer densities.  
Compared to the more commonly used coarse-grained polymer models with 
hard-core monomer interactions, the Laradji model has two advantages:
First, it can be simulated very efficiently, since the explicit evaluation
of pair interactions is often the most time consuming part in a simulation.
Second, it uses soft potentials, hence equilibration times are comparatively
short. The model does not account for packing effects and for topological
constraints (which may restrict the conformational phase space for strongly
adsorbed, quasi-two dimensional polymers). However, these are not in the focus
of the present study.

Specifically, the model system is a monodisperse brush in a good solvent
containing a single minority chain in a volume $V=L_{x}\cdot L_{y}\cdot L_{z}$.
We use periodic boundaries along $x$ and $y$ directions, while impenetrable
boundary walls are placed at $z=0$ and $z=L_{z}$. Polymer chains
are modelled by the discrete Gaussian bead-spring model with spring
constant $\frac{3k_{B}T}{2a^{2}}$, where $a$ is the statistical
bond length, $k_{B}$ is the Boltzmann constant, and $T$ the temperature.
We will use $a$ as the basic length unit, and $k_{B}T$ as the energy
unit. All chains are attached to a substrate placed at $z_{0}$ at
one end, where $z_{0}$ is chosen smaller than $a$. The impenetrable
wall at $z=0$ exerts an attractive potential to all minority chain
beads with strength $\varepsilon$ in a range of $0\leqslant z\leqslant a$.
There is no explicit solvent in the model
and all the relevant interactions in the good solvent case are represented
through an effective excluded volume potential between monomers (beads).
The Edwards type Hamiltonian of the system reads 
\begin{eqnarray}
\beta\mc H & = & \frac{3}{2a^{2}}\sum_{\alpha=1}^{n_{b}}\sum_{j=1}^{N_{b}-1}\big(\mb R_{\alpha j}-\mb R_{\alpha,j-1}\big)^{2}\nonumber \\
 & + & \frac{3}{2a^{2}}\sum_{j=1}^{N-1}\big(\mb R_{0j}-\mb R_{0,j-1}\big)^{2}\nonumber \\
 & + & \frac{v_\mathrm{bare}}{2}\int d\mb r\hat{\rho}_{t}^{2}(\mb r)+\int d\mb rU_{\mathrm{ads}}(\mb r)\hat{\rho}_{m}(\mb r),
\end{eqnarray}
where $\beta\equiv1/k_{B}T$, $\mb R_{\alpha j}$ denotes the position
of the $j$-th bead in the $\alpha$-th chain (the minority chain
has the index $\alpha=0$), $n_{b}$ is the total number of brush
chains, $\hat{\rho}_{m}$ denotes the local density of minority chain
monomers, $\hat{\rho}_{m}=\sum_{j}\delta(\mb r-\mb R_{0j})$, and
$\hat{\rho}_{t}\equiv\hat{\rho}_{b}+\hat{\rho}_{m}$ the total density
of monomers with $\hat{\rho}_{b}\equiv\sum_{\alpha=1}^{n_{b}}\sum_{j}\delta(\mb r-\mb R_{\alpha j})$ being the local density of brush monomers.
(Here and in the following, the subscript $b$ will be used to denote
the brush chains, and $m$ will denote the minority chain.) The first
two terms in the Hamiltonian describe
the Gaussian stretching energy (bonded interactions) between two neighboring
monomers in the same chain. The third term represents the non-bonded
effective interactions between polymer beads. For a good solvent the
excluded volume parameter $v_\mathrm{bare}$ is larger than zero. The fourth term
describes the adsorption between the substrate and the minority beads
with a step-like adsorption potential, 
\begin{equation}
U_{\mathrm{ads}}(\mb r)=\left\{ \begin{array}{cc}
-\varepsilon, & 0\leqslant z\leqslant a\\
0, & \mathrm{otherwise}
\end{array}\right.\label{eq:adsorption_potential}
\end{equation}

In the MC simulations, local densities are
extracted from the position of the beads by using the Particle-to-Mesh
technique \cite{Particle_Mesh}, which provides a way for the smoothing
of density operators. In the present work we use the zeroth order
scheme. The system is divided into cubic cells of size $b^{3}$ whose
centers define the grid points, and the local densities evaluated
by counting the total number of beads in the corresponding cells divided
by the cell volume, are defined at these grid points. The smoothed
monomer densities for the minority chain and brush chains are denoted
as $\rho_{m}$ and $\rho_{b}$, respectively. This implies that two
beads interact only when they are in the same cell, and
thus the size of the cell is indirectly linked to the interaction
strength of non-bonded interactions. Higher order schemes are conceivable,
but computationally more expensive. After the density operators are
smoothed over the cell volume, the Hamiltonian acquires the familiar
form used in the SCF theory \cite{Hybrid_PF}, hence SCF calculations
and MC simulations can be compared directly.

In the present study, the brush chain length
is fixed at $N_{b}=100$. In most MC of calculations
we choose the excluded volume parameter $v_\mathrm{bare}=1$ 
which would correspond to a Flory-Huggins
parameter $\chi=0$ in an explicit solvent solution (athermal solvent
condition), and the size of the averaging cell is taken $b=1$. We model
relatively dense monodisperse polymer brushes in good solvent with
the surface grafting density $\sigma=0.1,\ 0.2,\ 0.3$, with the corresponding
overlap parameters $\sigma R_{gb}^{2}\simeq2,\ 3,\ 5$, where $R_{gb}$
is the mean radius of gyration for an ideal brush polymer. The grafting
points of the chains on the substrate were fixed on a regular square
lattice.The system size was chosen $L_{x}=40$, $L_{y}=40$, and $L_{z}=100$
in most cases. To assess the influence of finite-size effects, we
have also carried out simulations of larger systems with size $L_{x}=60$,
$L_{y}=60$, and $L_{z}=100$ for selected parameter values. The results
were identical within the error.

MC simulations \cite{book_simulation} were
carried out according to the standard Metropolis criterion. At every
MC update, we try to move the position of one chosen monomer to a
new position with a distance in space comparable to the bond length.
This trial move results in an energy change including the bonded energy
and non-bonded energy, and it is accepted or rejected according to
the Boltzmann probability in the usual way. In all the simulations
$3\times10^{5}$ MC steps per monomer were performed to equilibrate
the system, and another $3\times10^{5}$ MC steps to extract statistical
averages. In order to get good statistics sampling for

the single minority
chain, an additional $10^{8}$ MC steps per monomer updating the minority
chain monomers were performed during the process of evaluating statistical
averages. Final statistical quantities were obtained by averaging
the results from 48 separate independent MC runs.

\begin{figure}[h]
\centering 
 \subfigure[]{ \label{Fig:pure_brush_s01} 
    \includegraphics[angle=0, width=7cm]{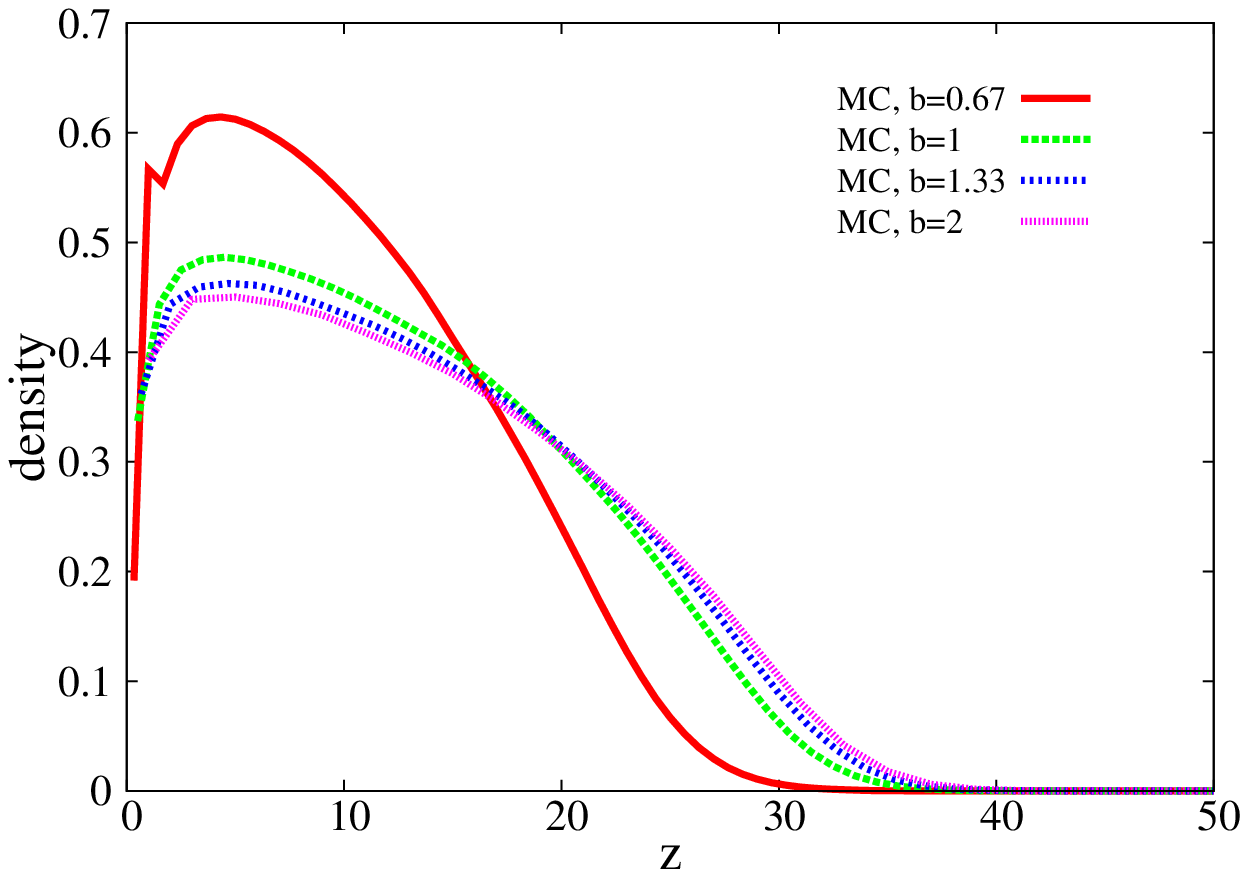}}
 \subfigure[]{ \label{Fig:pure_brush_s03} 
    \includegraphics[angle=0, width=7cm]{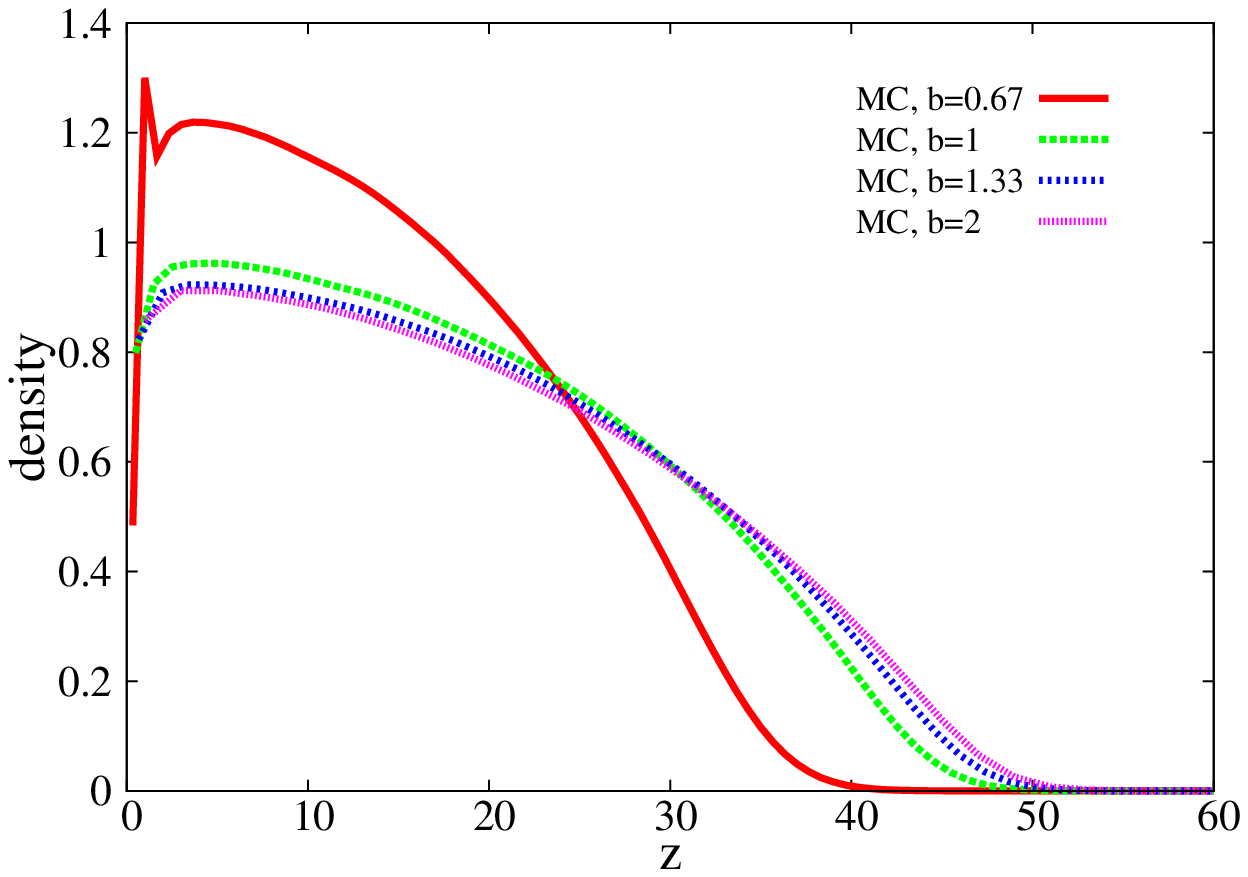}}
\caption{Monomer density profiles for pure brushes at grafting
density $\sigma=0.1$(a) and $\sigma=0.3$(b) from MC simulations
obtained with interaction cells of different size $b$ as indicated.}
\label{Fig:pure_brush} 
\end{figure}

\section{Homogenous brush and renormalization of the second
virial coefficient}\label{sec:brush}

We first analyze the simulation results for homogenous monodisperse brushes
which are very well understood.  In contrast to common MC polymer models our
model contains an additional free parameter, the cell size $b$, defining the
averaging volume for the local density operator, and the excluded volume
parameter $v_\mathrm{bare}$. Fig.\ \ref{Fig:pure_brush} shows the density
profiles of homogeneous brushes with two different grafting densities
$\sigma=0.1$ and $\sigma=0.3$, and various values of the averaging cell size
$b$. It is clear that the choice of cell size has a pronounced effect on the
brush density profile. To rationalize this effect we turn to the analytical SCF
theory developed over 20 years ago \cite{critical_adsorption,
Alexander,de_Gennes,Zhulina,Milner_analytical} and subsequently verified by
simulations and experiments.

Within the second virial approximation for good solvent conditions 
and assuming Gaussian elasticity for the chains, 
the density profile $\rho_{b}(z)$ has a parabolic form 
\begin{equation}
\rho_{b}(z)=\frac{3\pi^{2}}{8v_\mathrm{eff}N_{b}^{2}}(H^{2}-z^{2})
\end{equation}
where the brush height is $H=(4v_\mathrm{eff}\sigma/\pi^{2})^{1/3}N_{b}$, and
$v_\mathrm{eff}$ an effective interaction parameter. Gaussian
elasticity is an important ingredient in the analytical theory and the fact
that the MC simulation model is also based on a Hamiltonian with a Gaussian
term describing chain connectivity allows a more direct comparison. It is known
that corrections due to finite chain extensibility become important for dense
grafting, but this is outside the scope of the present paper. On the other hand,
modifications of the chain elasticity  due to the excluded volume effects are
naturally taken into account by our MC simulations. The mean-field potential
associated with the density, profile, $V_{b}(z)=v_\mathrm{eff}\rho_{b}(z)$ is
also parabolic 
\begin{equation}
V_{b}(z)=V_{0}-\frac{3\pi^{2}z^{2}}{8N_{b}^{2}}\label{eq:brush_potential}
\end{equation}
where $V_{0}=\frac{3}{2}(\pi v_\mathrm{eff}\sigma/2)^{2/3}$ is the potential
at the grafting surface. According to the theory (see for example
\cite{Milner_analytical}), the brush density profiles evaluated for
the same model at different values of $\sigma$ and $N_{b}$ should
collapse in the rescaled coordinates $\frac{\rho_{b}(z)}{\sigma^{2/3}}$
vs $\frac{z}{\sigma^{1/3}N_{b}}$ with one adjustable parameter 
related to the interaction parameter
$K=H/(\sigma^{1/3}N_{b})=(4 v_\mathrm{eff}/\pi^{2})^{1/3}$
\begin{equation}\label{fitting_B}
\frac{\rho(z)}{\sigma^{2/3}}=\frac{3}{2K^{3}}
\left[K^{2}-\left(\frac{z}{\sigma^{1/3}N_{b}}\right)^{2}\right]
\label{eq:collapse}
\end{equation}

\begin{figure}[h]
\centering 
 \subfigure[]{ 
    \label{Fig:density_scaled} 
    \includegraphics[angle=0, width=7cm]{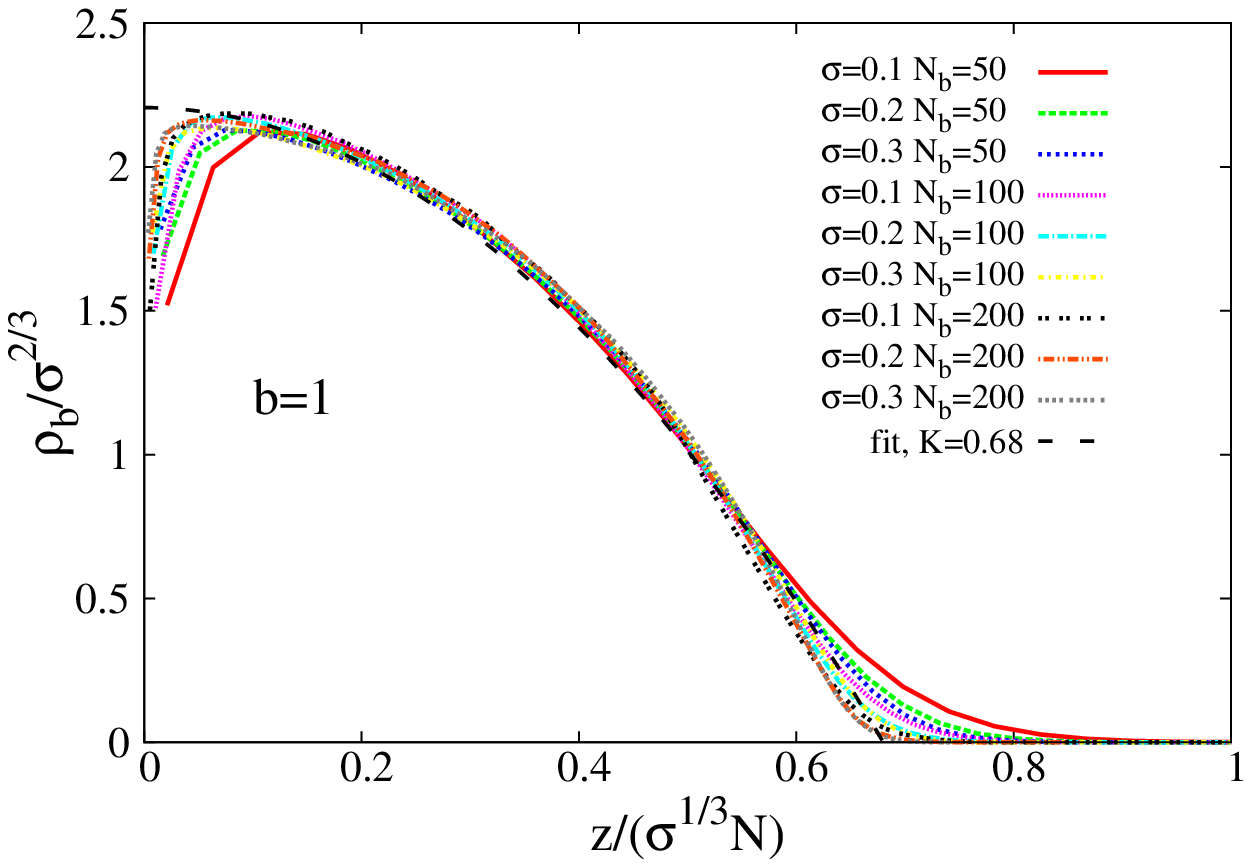}}
 \subfigure[]{ \label{Fig:v_b} 
    \includegraphics[angle=0, width=7cm]{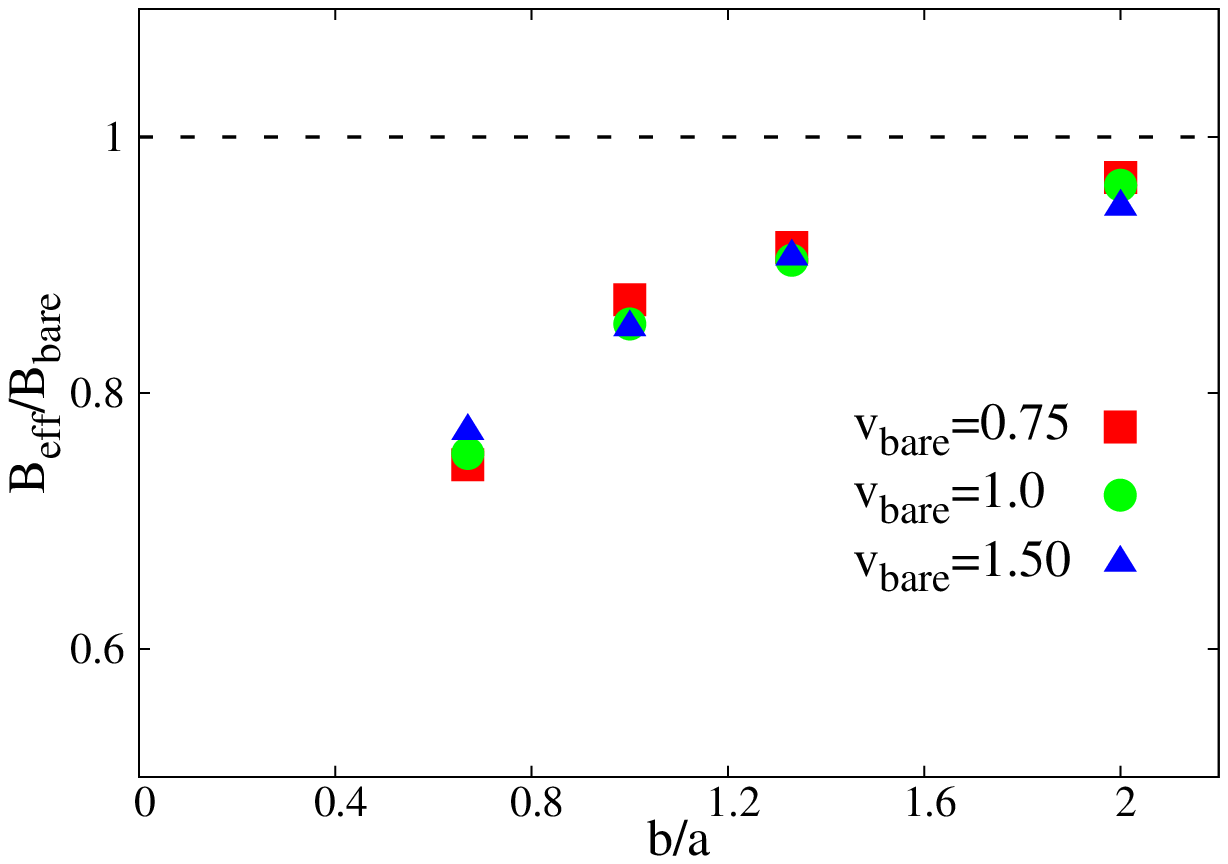}}
\caption{Density profile of polymer brushes (a) in rescaled coordinates
$\rho_b(z)/\sigma^{2/3}$ vs $z/(\sigma^{1/3}N_b)$ calculated by MC simulations
with $v_\mathrm{bare}=1$ for $N_b=50, 100, 200$  and different grafting
densities $\sigma=0.1,\ 0.2,\ 0.3$. Dashed line shows the best fit to Eq.\
(\protect\ref{fitting_B}), corresponding to  $K=0.68$. Panel (b) shows the
dimensionless ratio $B_\mathrm{eff}/B_\mathrm{bare}$ as a function of $b/a$
with $B = \frac{1}{2} b^3 (1-{\rm e}^{-v/b^3})$ for different $v_\mathrm{bare}$.
These points are calculated from $K$ obtained from the best fitting curves at
fixed chain length $N_b=100$.} 
\label{Fig:scale_brush} 
\end{figure}

The rescaled MC density profiles with $\sigma=0.1,\ 0.2,\ 0.3$ and
$N_{b}=50,\:100,\:200$ are presented in Fig.\ \ref{Fig:density_scaled}, for a
specific choice of the averaging cell size, $b=1$ and the ``bare'' excluded
volume parameter $v_\mathrm{bare}=1$ . The quality of the collapse is good, and
one can extract the value of the effective interaction parameter
$v_\mathrm{eff}=0.776$, which differs from $v_\mathrm{bare}$. The same procedure
was repeated for several other values of the $b$ parameter, and different bare
excluded volume parameters $v_\mathrm{bare}$.  
To evaluate the difference between $v_{\mathrm{eff}}$ and
$v_{\mathrm{bare}}$, we calculate the corresponding exact second exact virial
coefficient in the simulation model, taking into account the finite
grid size $b$. For given interaction parameter $v$, it is given by 
$B = \frac{1}{2} b^3 (1-\mathrm{e}^{-v/b^3})$. The ratio 
$B_{\mathrm{eff}}/B_{\mathrm{bare}}$ for the best fit values of
$v_{\mathrm{eff}}$ are plotted as a function of $\frac{b}{a}$ in
Fig.\ \ref{Fig:v_b}. The data for different values of $v_{\mathrm{bare}}$
collapse nicely.

We conclude that the renormalization
effect can be presented in terms of a dimensionless ratio
$\frac{B_\mathrm{eff}}{B_\mathrm{bare}}$, which is a universal function of a
single dimensionless parameter $\frac{b}{a}$ (or, equivalently, $b
\sqrt{N}/R_g$).  Universality in this context does not mean a broad class of
different models but rather that other model parameters, such as $\sigma$ and
$N_{b}$ do not enter explicitly. It is clear from Fig.\ \ref{Fig:v_b} that at
cell size $b/a=2$ or larger, the curve saturates and $v_\mathrm{eff}$
approaches the bare value. 
Hence the system approaches mean-field behavior if the interaction 
range $b$ of the monomers becomes large, as one would expect. At smaller
$b/a$, local monomer correlations become important and renormalize
the effective interaction parameter. Renormalization of monomer interaction
parameters has also been observed in Edwards-type simulations of polymer melts 
\cite{Particle_Mesh}.

\section{Minority chain in a brush: theoretical background}\label{sec:theory}

The theory was briefly sketched in Ref.\ \cite{Klushin_switch}. In this
section, we present it in more detail. We consider a relatively
dense monodisperse polymer brush in a good solvent containing a single
surface-active minority chain. The length of the minority chain is denoted as
$N$, while the length of the brush chain is denoted as $N_b$. The minority
chain and the brush chains are chemically different and have different
interactions with the substrate, i.e., the substrate adsorbs minority chain
monomers with strength $\varepsilon$, while it is neutral to the brush chain
monomers. Both the minority and the majority chains are flexible with the same
statistical Kuhn length $a$ which is taken as the unit length; the excluded
volume parameter is also the same for both chain types and is positive
corresponding to good solvent conditions. This is 
fully consistent with the MC model.

We treat the inter- and intrachain interactions in the mean-field approximation
and neglect the back effect of the change in the minority chain conformation on
the surrounding brush. Hence, the conformation of a minority chain is affected
by a fixed mean-field potential profile consisting of the repulsive
contribution determined by the brush density and a short-ranged attraction due
to the solid substrate. The minority chain itself is described by an ideal
continuum model, since intrachain excluded volume effects are screened out
considerably within the brush thickness.

The analytical description of the minority polymer chain is based on the
continuum approach. The Green's function $G(z,N)$, i.e. the total statistical
weight of the minority chain grafted at the substrate ($z_0=0$) as a function
of the free end position, $z$, is a solution of the Edwards equation
\begin{equation}
 \frac{\partial G(z,s)}{\partial s}=\frac{1}{6}\frac{\partial^2G(z,s)}{\partial z^2}-V(z)G(z,s)
\end{equation}
taken at $s=N$, with the initial condition $G(z,0)=\delta(z)$. The total
potential $V(z)=V_b(z)+V_\mathrm{ads}(z)$, where $V_\mathrm{ads}(z)$ is the
adsorption attraction potential, and $V_b$ is the mean-field brush potential
given by Eq.\ (\ref{eq:brush_potential}). The adsorption potential
$V_\mathrm{ads}(z)$ is described in the analytical SCF approach as an
attractive pseudopotential $V_\mathrm{ads}(z)\propto-c\delta(z)$ where $c$ is the
adsorption interaction parameter in the continual model. This parameter was
introduced by de Gennes \cite{de_Gennes_boundary} to replace the real
adsorption potential through the boundary condition
$[G(z,s)]^{-1}\frac{\partial G(z,s)}{\partial z}\big|_{z=0}=-c$. Close to the
substrate, $z\ll H$, the brush potential changes very little, $V_b(z)\simeq
V_0=\kappa\sigma^{2/3}$ where $\kappa=\frac{3}{2}(\pi
v_\mathrm{eff}/2)^{2/3}\simeq 1.71$ is a shorthand notation for the numerical
coefficient

The adsorbed state is thus approximately described by the standard Green's
function modified by the potential $V_0$ \cite{brush_minority}
\begin{equation} 
G_\mathrm{ads}(z)=2ce^{N(-V_0+c^2/6)-cz}.
\end{equation}
Integration over $z$ gives the partition function
\begin{equation}\label{eq:partition_ads} 
Q_\mathrm{ads}=2e^{N(-V_0+c^2/6)}
\end{equation}
In order to connect this expression to the MC model we note that $c^2/6$ has
the meaning of the negative free energy of adsorption per segment (the chemical
potential) in an asymptotically long chain, i.e., $\mu=c^2/6$. When comparing
the actual brush density profile with the theoretical parabolic shape, one
concludes that the pseudopotential must account for the combined effect of the
actual adsorption potential (a step of unit width with energy $\varepsilon<0$)
and the short-range depression in the brush density near the wall. It is known
that for weak adsorption, the free energy per monomer is determined by the
crossover exponent $\phi:\mu\sim(\varepsilon-\varepsilon_c)^{1/\phi}$. For an
ideal chain, $\phi=1/2$ while the value for a chain with excluded volume was a
subject of extensive investigations and prolonged debates \cite{Grassberger, Klushin_adsorption}. 
For any practical purposes, the adsorption of relatively short chains
is very accurately described by the ideal value $\phi=1/2$. Here, for weak
adsorption, we adopt the expression
\begin{equation}\label{eq:mu_epsilon}
\mu(\varepsilon)=\alpha(\varepsilon-\varepsilon_c)^2 
\end{equation}
where $\alpha$ and $\varepsilon_c$ are model-dependent constants to be
extracted from the MC data. It is also known that at stronger adsorption, the
free energy deviates from Eq.\ (\ref{eq:mu_epsilon}), and for very large
$\varepsilon$, it can simply be estimated as $\mu=\varepsilon-\mathrm{const}$,
where the constant shift is given by the limiting entropy difference per
monomer between the coil and the fully adsorbed state.

The state with the free end exposed at the brush edge has no contacts with the
surface. Hence the Green's function is given by the known solution of the
Edwards equation for a purely parabolic potential and a neutral solid surface
\cite{brush_minority_1997}
\begin{equation}\label{eq:Green_exposed}
G_\mathrm{ex}(z,N)=\frac{\pi}{2}\left[\frac{3}{N_b\sin\Big(\frac{\pi
N}{2N_b}\Big)}\right]^{3/2}ze^{-\frac{3\pi}{4N_b}\cot\big(\frac{\pi
N}{2N_b}\big)z^2}.
\end{equation}
For longer minority chains, $N>N_b$, the Green's function of
Eq.\ (\ref{eq:Green_exposed}) shows unlimited monotonic increase with $z$ since
the solution refers to a potential extending to infinity and to an infinitely
extendable Gaussin chain. In a realistic situation the minority chain does not
make excursions well beyond $z=H$. To account for this fact we truncate
$G_\mathrm{ex}(z,N)$ at $z=H$. If minority chains are just slightly longer
$\Delta\equiv N-N_b\ll N_b$, (which is the situation of interest in this paper), 
the Green's function simplifies to
\begin{equation}\label{eq:Green_ex}
G_\mathrm{ex}(z,N)=\left\{\begin{array}{cc}
\frac{\pi}{2}\Big(\frac{3}{N_b}\Big)^{3/2}ze^{V_0\Delta(\frac{z}{H})^2}, &
z\leqslant H \\ 0, & z>H \end{array}\right..
\end{equation}
This gives the partition function
\begin{equation}\label{eq:partition_ex}
Q_\mathrm{ex}=\frac{2}{\pi\Delta}(3N_b)^{1/2}e^{V_0\Delta}
\end{equation}
assuming $e^{V_0\Delta}\gg 1$. Eqs (\ref{eq:partition_ads}) and
(\ref{eq:partition_ex}) clearly show that in the limit of $N_b\to\infty$,
$\sigma=\mathrm{const}$, $\frac{N-N_b}{N_b}=\mathrm{const}$ the free energies
of the two states are extensive and switching become a classical first-order
phase transition.

Based on this analytical approach, the transition properties between the
adsorbed state and exposed state can be investigated. For example, one could
define the transition point as the condition
$Q_\mathrm{ads}(c^*)=Q_\mathrm{ex}$. Omitting logarithmic correlations this
gives $\mu^*=V_0\frac{\Delta}{N}$. Under the condition $U\frac{\Delta}{N}\ll 1$
one can use the asymptotic expression Eq.\ (\ref{eq:mu_epsilon}) to obtain the
adsorption energy at the transition
\begin{equation}\label{eq:transition_point}
-\varepsilon^*=-\varepsilon_c+(\kappa/\alpha)^{1/2}\sigma^{1/3}\Big(\frac{\Delta}{N}\Big)^{1/2}
\end{equation}
For larger grafting densities and chain length differences one expects
deviation from the scaling form of of Eq.\ (\ref{eq:transition_point}), since
the quadratic dependence of $\mu(\varepsilon)$ may have a limited range. In the
extreme case of a very long minority chain, $N\gg N_b$, the transition point
satisfies the condition $\mu^*=V_0$. Since the mean-field potential $V_0$ is in
the range $V_0\sim0.36-0.8$, the limiting value of $\varepsilon^*$ remains of
order 1.

The sharpness of the transition is characterized by the width of the adsorption
parameter change, $\delta\varepsilon$, sufficient to produce a reliable
switching from the adsorbed state with the average distance of the free end of
the minority chain $Z_\mathrm{end}\ll H$ to the exposed state with
$Z_\mathrm{end}\simeq H$: $\delta\varepsilon =
(\frac{1}{H}\frac{dZ_\mathrm{end}}{d\varepsilon})^{-1}$ where the derivative is
evaluated at the transition point itself. A standard recipe for a two-state
model \cite{two_state_model} gives $\delta\varepsilon=(\frac{1}{4}\frac{d\ln
Q_\mathrm{ads}}{d\varepsilon})^{-1}\big|_{\varepsilon=\varepsilon^*})^{-1}$
which combined with Eq.\ (\ref{eq:partition_ads}) results in
\begin{equation}\label{eq:sharpness scaling}
\delta\varepsilon=\frac{2}{\kappa}(\kappa/\alpha)^{1/2}\sigma^{-1/3}(N\Delta)^{-1/2}
\end{equation}

Finally, the switching time is essentially determined by the barrier separating
the two relevant states. Hence we estimate the barrier height at the transition
point, $U_\mathrm{barrier}$, in terms of the change in the Green's function of
the exposed state,
$U_\mathrm{barrier}=\ln\frac{G_\mathrm{ex}(z=H)}{G_\mathrm{ex}(z=z_\mathrm{barrier})}$.
It follows from Eq.\ (\ref{eq:Green_ex}) that up to a numerical coefficient
close to unity
\begin{equation}\label{eq:barrier scaling}
U_\mathrm{barrier}\sim V_0\Delta\sim\kappa\sigma^{2/3}\Delta
\end{equation}

\section{Simulation Results and discussion}
\label{sec:results}

\subsection{Adsorbed and exposed conformations of the minority chain}

\begin{figure}[h]
  \centering
  \subfigure[]{
    \label{Fig:snapshot_adsorbed} 
    \includegraphics[angle=0, width=7cm]{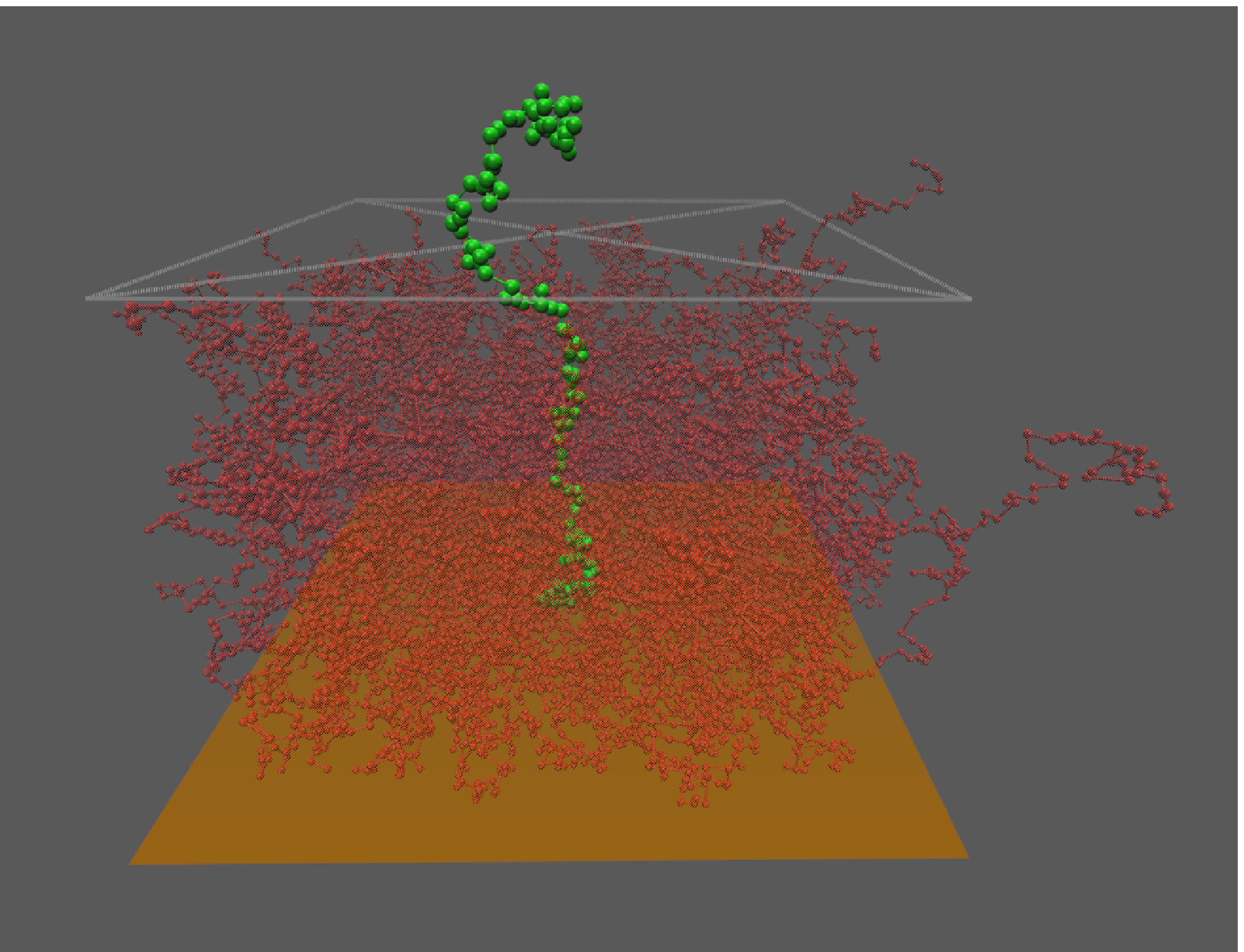}}
  \subfigure[]{
    \label{Fig:snapshot_exposed} 
    \includegraphics[angle=0, width=7cm]{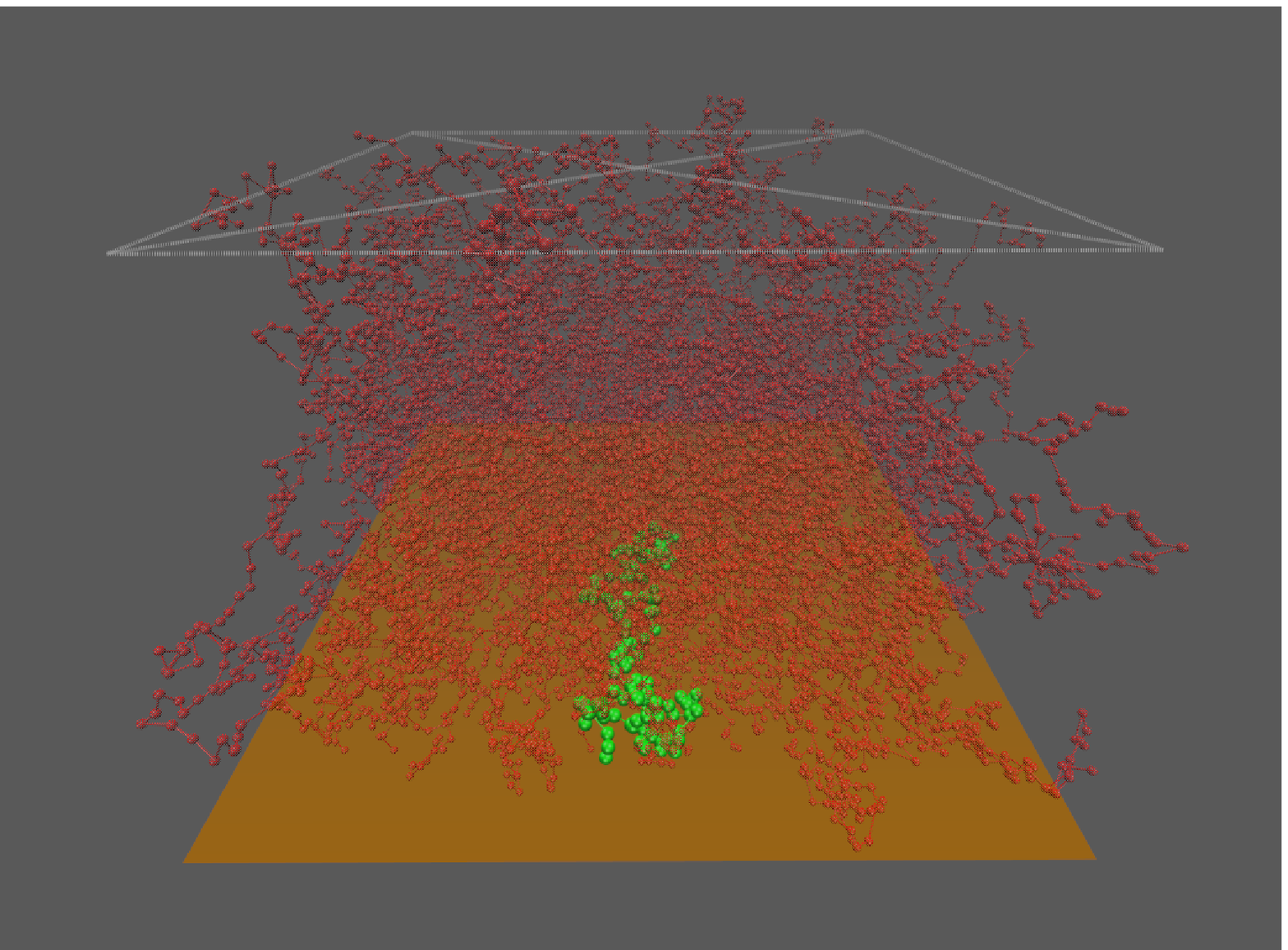}}
  \caption{
Simulation snapshots of a polymer brush with $N_b=100$ and grafting density
$\sigma = 0.1$ containing a minority chain with $N=120$ in the exposed state (a)
at $\varepsilon=0$ and the adsorbed state (b) at $\varepsilon=-0.7$. The minority chain beads
are colored green, while the brush chain beads are colored red and made
transparent. Brush height $H$ is shown by empty square above for orientation.
}
  \label{fig:snapshots} 
\end{figure}

Fig.\ \ref{fig:snapshots} shows typical simulation snapshots of a brush
containing a minority chain $N=120$ (monomers) in an exposed and an adsorbed
state. There is a substrate surface below and a virtual surface above showing
the brush height $H$. As can be seen a small fraction of monomer units are
located above $H$. These monomers form a tenuous brush exterior. The minority
chain is shown in green color. It is longer than the majority brush chains with
$N_{b}=100$. We characterize the length of the minority chain by a parameter
$\Delta=N-N_{b}$ (In the snapshot, $\Delta$ is given by $\Delta=20$).  The
minority chain in the exposed state has a stem-crown-like conformation.

\begin{figure}[h]
  \centering
  \subfigure[]{
    \label{Fig:density_N_e0} 
    \includegraphics[angle=0, width=7.0cm]{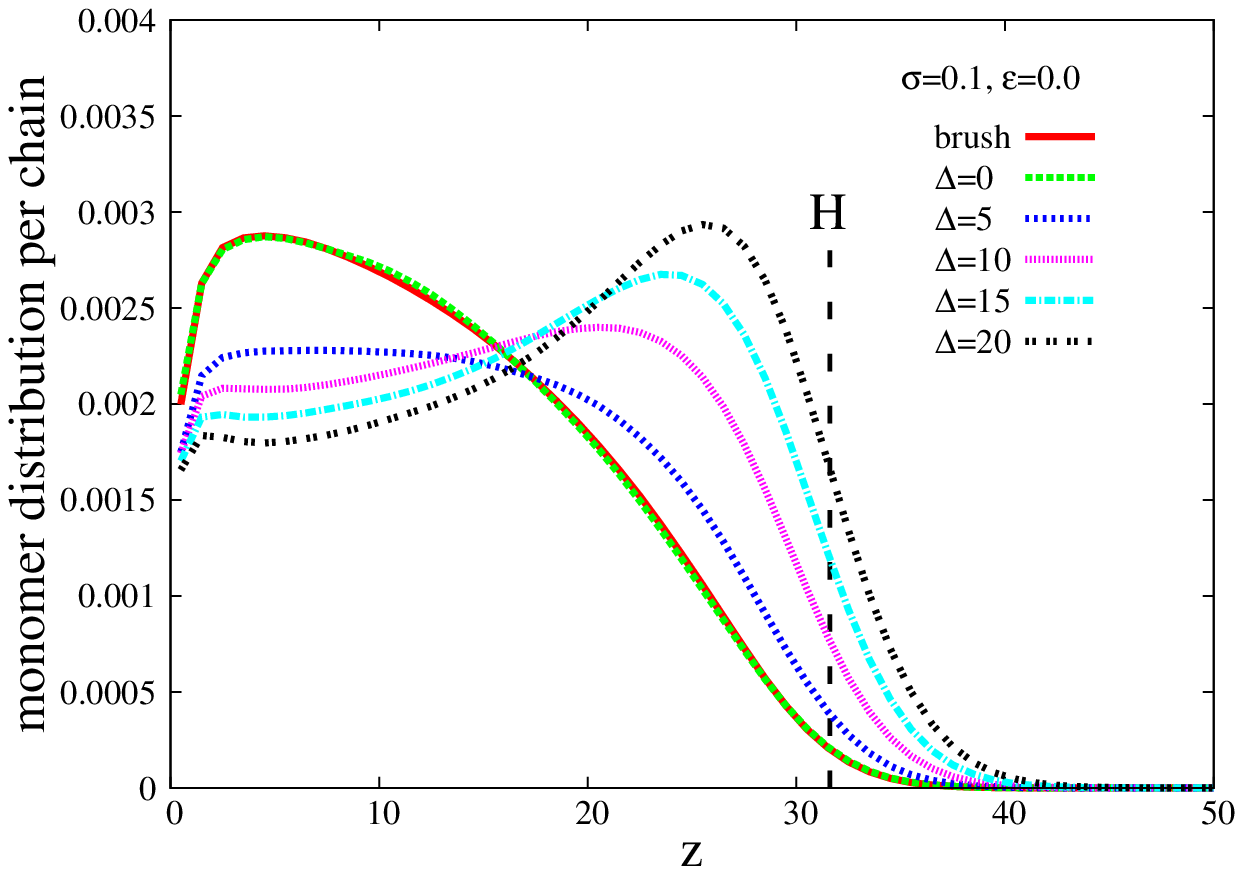}}
  \subfigure[]{
    \label{Fig:density_N_e0.7} 
    \includegraphics[angle=0, width=7.0cm]{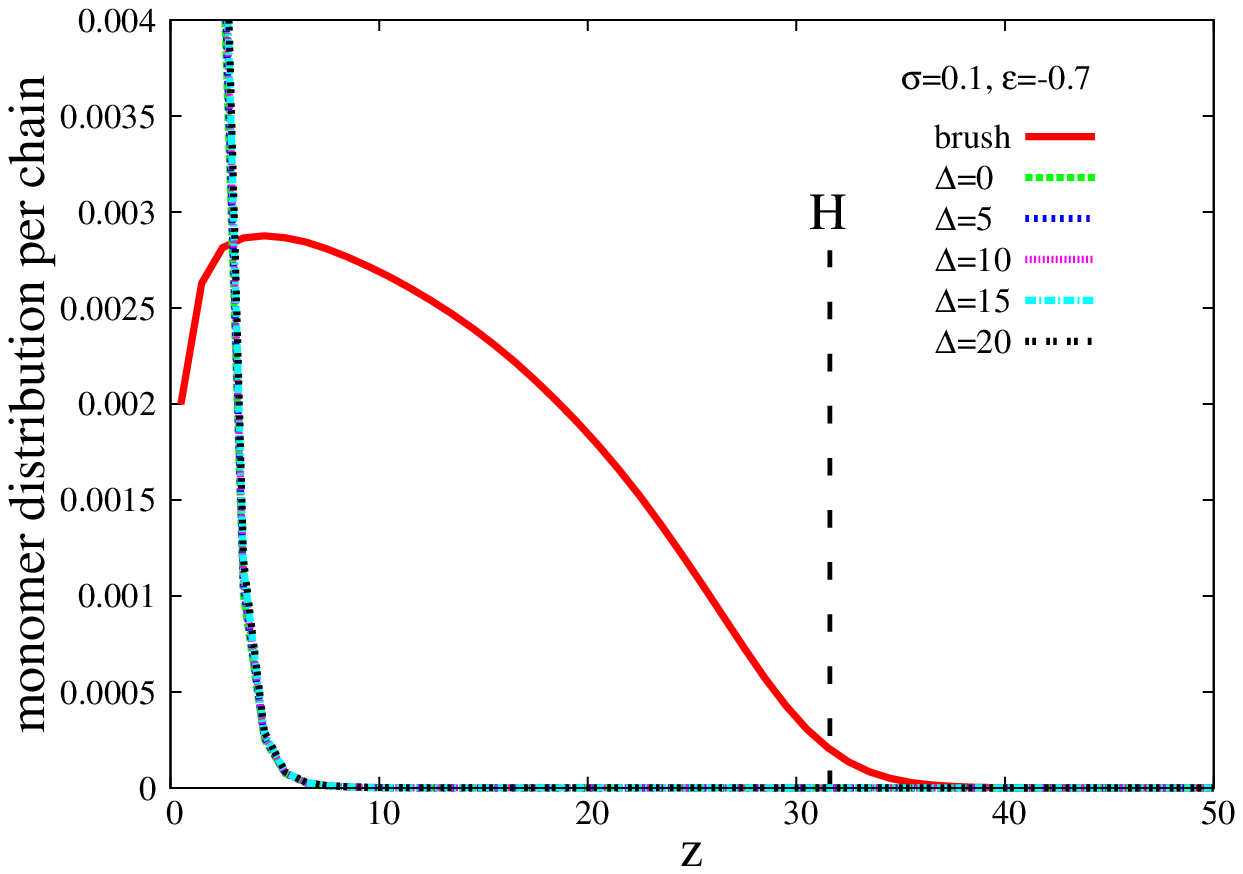}}
  \caption{
Longitudinal monomer density profiles of minority chains of different length
$N=N_{b}+\Delta$ in the exposed state at $\varepsilon=0.0$ (a), and in the
adsorbed state at $\varepsilon=-0.7$ (b) as obtained from MC simulations. The
brush parameters are $N_{b}=100$ at $\sigma=0.1$.  The density profile of the
brush chain normalised to unity with is shown for comparison by solid red line.
Brush height $H \approx 32$ is shown by vertical dashed line.
}
  \label{Fig:density_z} 
\end{figure}

The main features of the snapshot can be also seen in Fig.\ \ref{Fig:density_z}
where the longitudinal monomer density profiles are shown for minority chains
with different $\Delta$ in the exposed state at $\varepsilon=0$ (panel a), and
in the adsorbed state at $\varepsilon=-0.7$(panel b). The density profile of a
homogeneous brush is also shown, and for comparison both profiles are
normalized in the same way. It is clear that at $\varepsilon=0$ and $\Delta=0$
the minority chain is identical to brush chains. For longer minority chains,
$\Delta>0$, the density profile deviates from a parabolic, monotonically
decreasing form (apart from a narrow depletion layer near the substrate) shape.
It is well understood that the standard parabolic shape is due to a specific
very broad end-monomer distribution. ``Partial'' density profiles produced by
chains with a given end position are initially increasing with $z$ with a
subsequent sharp drop. Only after averaging over end positions one obtains the
actual density profile. The most pronounced effect is observed for the longest
minority chain with $\Delta=20$.  The profile can be understood if we take
chain conformations with the end-monomer positioned on average close to $z=H$
(shown by the vertical dashed line) with a typical fluctuation of a few monomer
lengths ($\sim5).$ For shorter minority chains the average end-hight is below
$H$ and the fluctuations are relatively larger, which makes the difference to
the majority chains less dramatic. In the strongly adsorbed state the density
profile exponentially decreases with $z$ and is insensitive to $\Delta$, see
Fig.\ \ref{Fig:density_N_e0.7}.

\begin{figure}[ht]
\centering
\includegraphics[angle=0,scale=0.6,draft=false]{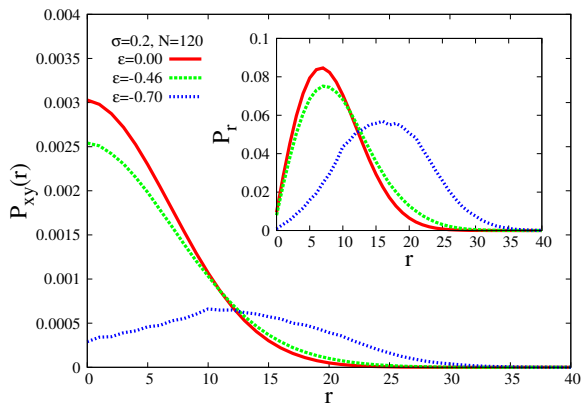}
\caption{
Lateral distribution of the free end of the minority chain as a function of
radial (in-plane) distance from the grafting point $r$ for $\sigma=0.2$,
$N=120$, and different adsorption strengths as indicated. The inset shows the
lateral probability given by $P_r=2\pi rP_{xy}(r)\Delta r$, where $\Delta r$=1.
The data are obtained by MC simulations
}
\label{Fig:lateral_probability} 
\end{figure}

Even though the minority chains are highly dilute, they perturb the brush
locally. Furthermore, the minority chain itself of course has lateral
structure. Fig.\ \ref{Fig:lateral_probability} shows the lateral distribution
of the end monomer $P_{xy}(r)$ as a function of the in-plane radial distance
$r=\sqrt{x^{2}+y^{2}}$ from the grafting point for different adsorption
energies $\varepsilon$ at grafting density $\sigma=0.2$ and chain length
$N=120$. Here $P_{xy}(r)$ was obtained by counting how often the minority chain
end was located in a cylindrical shell ranging from $r$ to $r+\Delta r$ and
dividing this by the volume of the shell and the total sampling number. The
thickness of the shell was chosen as $\Delta r=1$. The results for
$\varepsilon=0$ in Fig.\ \ref{Fig:lateral_probability} show that $P_{xy}(r)$
exhibits Gaussian behavior for nonadsorbing chains. With increasing adsorption
strength, the distribution becomes broader, and a depletion region develops
close to $r=0$: Chain ends are pushed away from the area close to the grafting
point, due to the fact that this area is already covered by adsorbed monomers.
This is also shown by the radial distribution density $P_{r}(r)=2\pi
rP_{xy}(r)$, which gives the probability to find the free end of the minority
chain located at distances ranging from $r$ to $r+\Delta r$ with $\Delta r=1$.
At strong enough adsorption the maximum of $P_{r}(r)$ moves towards larger
values of $r$ since the flattened adsorbed state dominates (Fig.\
\ref{Fig:lateral_probability}, inset).

\begin{figure}[h]
\centerline{\includegraphics[angle=0,scale=0.6,draft=false]{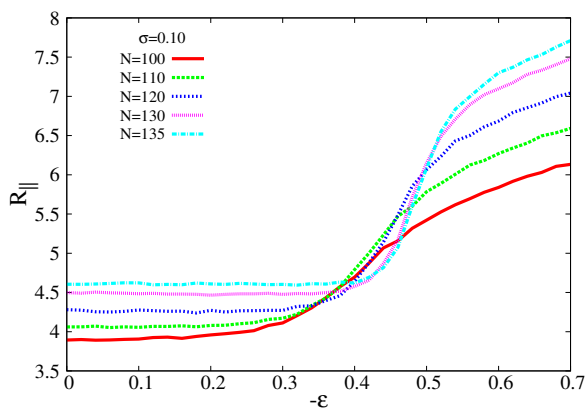}}
\caption{
Lateral size of the minority chain $R_\|$ defined as
$R_\|=\sqrt{R^2_{gx}+R^2_{gy}}$ vs the adsorption strength for $\sigma=0.10$
and different minority chain lengths. The data are obtained by MC simulations
}
\label{Fig:lateral_size} 
\end{figure}

To measure more quantitatively how the minority chain extends in the lateral
direction, we calculate the parallel root-mean-squared radius of gyration
defined as $R_{\|}=\sqrt{R_{gx}^{2}+R_{gy}^{2}}$, where $R_{gx}$ and $R_{gy}$
are the mean-square radius of gyration of the minority chain in the $x$ and $y$
directions, separately.

\begin{figure}[h]
\centerline{\includegraphics[angle=0,scale=0.6,draft=false]{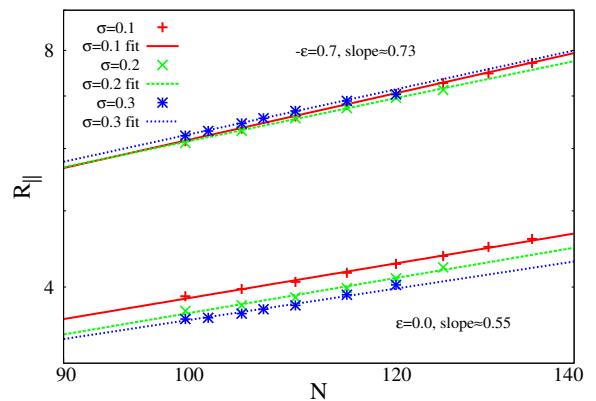}}
\caption{
Lateral size of the minority chain vs chain length for different grafting
density in the exposed state ($\varepsilon=0$) and in a strongly adsorbed state
$-\varepsilon=0.7$. The symbols denote the results from MC simulations, lines 
are obtained by a linear fitting after taking the logarithm of the original 
data
}
\label{Fig:Rxyg} 
\end{figure}

Fig.\ \ref{Fig:lateral_size} shows $R_{\|}$ as a function of the adsorption
strength at $\sigma=0.1$ for different minority chain lengths. From these
curves it can be seen that the lateral size of the minority chain remains
almost unchanged up to $\varepsilon\sim-0.4$, and then starts to increase,
which is another signature of the conformational transition from the desorbed
to the adsorbed state.  Fig.\ \ref{Fig:Rxyg} shows the lateral size of the
minority chain in the exposed state with $\varepsilon=0$ and in a strongly
adsorbed state with $-\varepsilon=0.7$, as a function of the minority chain
length for several grafting densities.  In both the exposed state and the
adsorbed state, $R_{\|}(N)$ is consistent with a power law. The best fit for
the exposed state ($\varepsilon=0$) gives $R_{\|}\propto N^{\nu_{e}}$, with
$\nu_{e}\thickapprox0.55$, which is larger than the Flory exponent expected for
ideal chains ($\nu=0.5$), and smaller than that expected for free chains in a
good solvent ($\nu=0.588$). This can be understood since part of the chain well
within the brush experiences semi-dilute conditions with screened excluded
volume interactions while the part near the tenuous brush exterior is swollen.
As one increases the grafting density at fixed length, the lateral size of the
exposed chain decreases. This has two reasons: First, the brush 
thickens with increasing grafting density, and consequently the fraction of 
swollen minority chain part outside the brush (the crown) decreases. Second,
the monomer density inside the (semi-dilute) brush increases, which
leads to a stronger screening of excluded volume interactions and a
decrease of lateral swelling of the inner part of the minority chain 
(the stem).  According to Fig.\ \ref{Fig:lateral_size}, the shrinking does 
not depend noticeably on the chain length $\Delta=N-N_b$ between minority 
chain and brush chains, i.e.\ , on the size of the crown. Hence we
conclude that the second effect is probably dominant.

For strongly adsorbed chains ($-\varepsilon=0.7$), we find a
relation $R_{\|}\propto N^{\nu_{a}}$, with $\nu_{a}\thickapprox0.73$, which is
close to the Flory exponent for two dimensional self-avoiding chains
($\nu=0.75$). In the strongly adsorbed state the lateral extension is almost
insensitive to the grafting density, see Fig.\ \ref{Fig:Rxyg}, upper line.
From the scaling point of view the adsorption blob size is quite small and thus
screening of the excluded volume interactions by other chains is ineffective.

\subsection{Transition properties}

\begin{figure}[h]
  \centering
  \subfigure[]{
    \label{Fig:Zend_MC} 
    \includegraphics[angle=0, width=7.0cm]{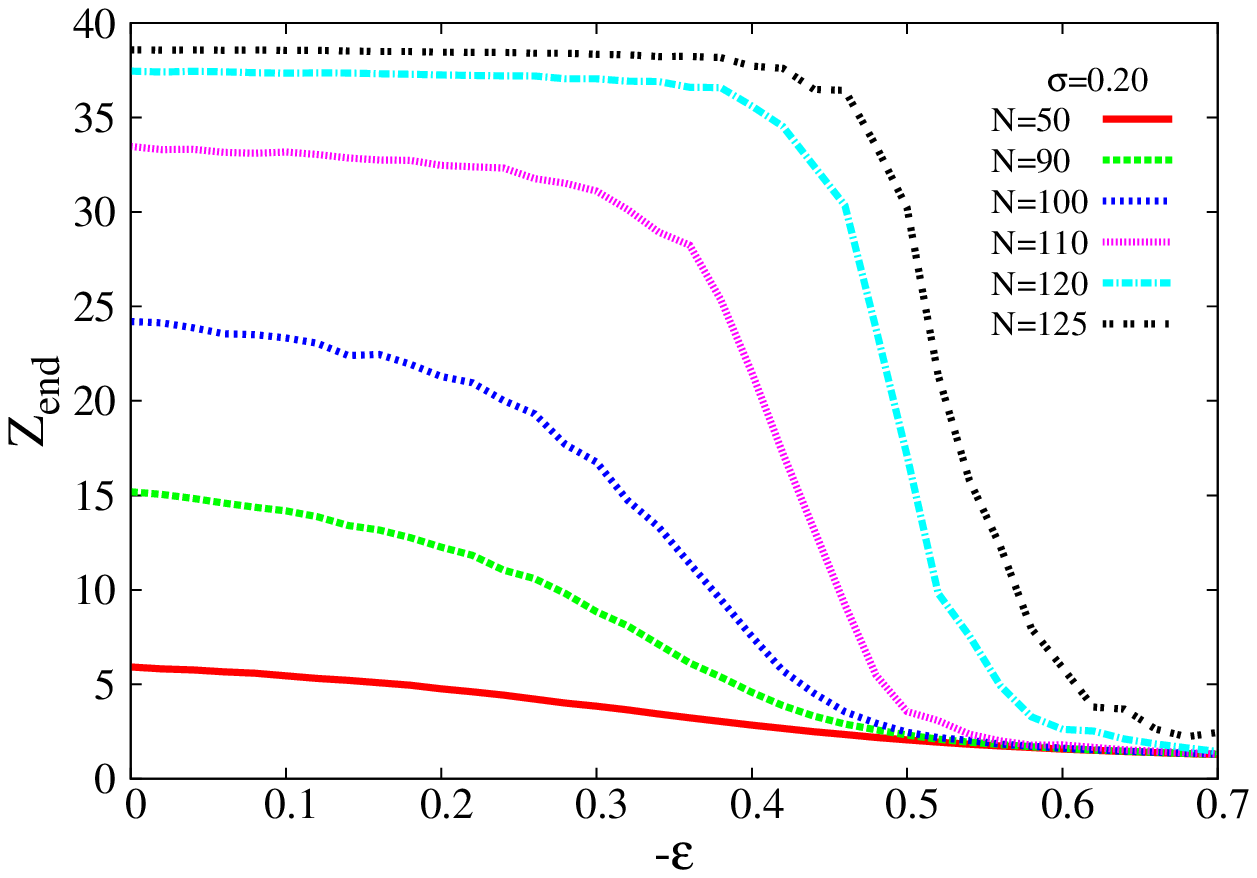}}
  \subfigure[]{
    \label{Fig:Zend_MC_SCF} 
    \includegraphics[angle=0, width=7.0cm]{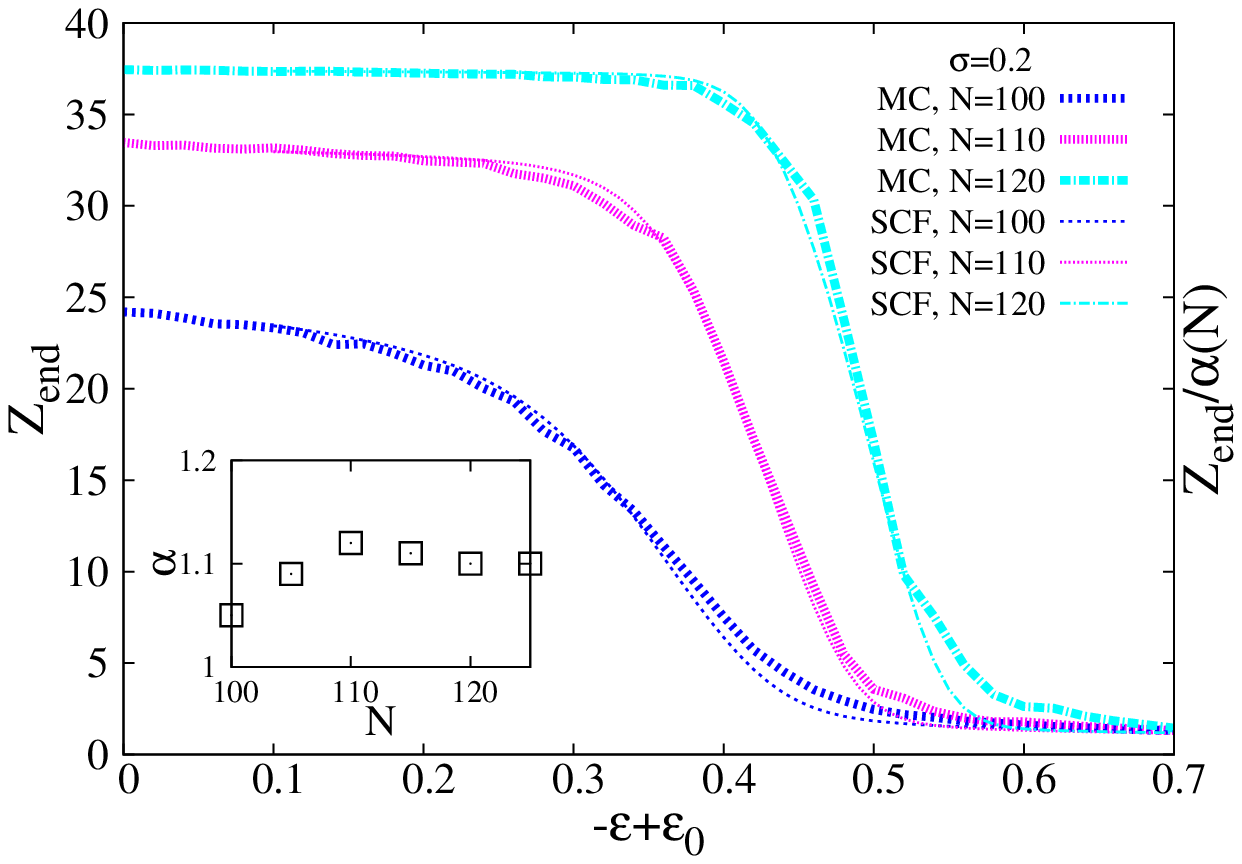}}
  \caption{
Average distance $Z_\mathrm{end}$ of the free end of the minority chain from
the surface vs adsorption strength for different chain lengths with grafting
density $\sigma=0.2$ obtained from MC simulations (a). Superposition of a
subset of these curves (MC) with the corresponding curves obtained by 
3d SCF calculations (SCF), where the SCF curves have been shifted by 
$\varepsilon_0=0.1$ and rescaled by an empirical factor $\alpha(N)$, 
given in the inset.  The SCF calculations were done with the renormalized 
interaction parameter $v_\mathrm{eff}$ extracted from 
Fig.\ \protect\ref{Fig:v_b}.
}
  \label{Fig:Zend} 
\end{figure}

Fig.\ \ref{Fig:Zend_MC} presents the average distance $Z_{\mathrm{end}}$ from
the free end of the adsorption-active minority chains to the solid substrate as
a function of the adsorption strength $-\varepsilon$, for minority chains of
several lengths in a brush with $N_{b}=100$ at $\sigma=0.2$. The two lower
curves represent the behavior of shorter minority chains with $N=50$ and
$N=90$.  In the desorbed state, short minority chains exist in a coil-like
conformation and hence the adsorption transition is expected to be continuous.
The curves are further smeared out due to the finite length of the chain.
In contrast, even a small positive increment $\Delta=N-N_{b}$ in the length of
minority chain results in a well-pronounced sharp transition from an exposed
state in which the free end of the minority chain is localized at the outer
space of the brush to an adsorbed state where the chain is localized close to
the substrate as illustrated in Fig.\ \ref{Fig:density_z}

Fig.\ \ref{Fig:Zend_MC_SCF} compares the MC results with the prediction of the
three-dimensional SCF, using the interaction parameter $v_\mathrm{eff}$
extracted from the MC simulations of pure brushes (Fig.\ \ref{Fig:v_b}).  The
SCF results differ from the MC date in two important aspects. First, they do no
capture the attractive contact interaction of the minority chain with the
surface accurately. This is not surprising, given that the interaction vanishes
on the scale of one grid size, and it leads to a shift of the adsorption curves
by a constant value $\varepsilon_0\approx 0.1$. Second, the density
fluctuations at the brush surface (see Fig.\ \ref{Fig:density_scaled})
effectively reduce the (spatial) range of the repulsive potential, hence the
minority chain is less stretched in the exposed state. This effect is even
observed for minority chains that have the same length than the brush chains,
$N=N_b$. We will see below that this leads to a reduction of the free energy
barrier between the exposed and the adsorbed state.  After shifting the
SCF curves by $\varepsilon_0$ and rescaling them with a factor $\alpha(N)$ which
was obtained empirically, they are in very good agreement with the MC results.

Curves such as those shown in Fig.\ \ref{Fig:Zend}, which give the average
distance of the minority chain free end from the surface as a function of the
adsorption strength, can be used to localize the crossover between the adsorbed
and the desorbed state. The transition point $-\varepsilon^{*}$ is then defined
as the point where the slope of the curve is maximal (for $\Delta > 0$).
The value of the slope at this point is used to extract the width of the
transition, $\delta\varepsilon$. This is done as follows. First we find the
point where $Z_{\mathrm{end}}$ vs $-\varepsilon$ has the maximum slope, and
calculate this slope as $l$. Then we draw a line through this maximum slope
point with the slope $l$ and find its intersection with (i) the abscissa and
(ii) with a line parallel to the abscissa that corresponds to the value of
$Z_{\mathrm{end}}$ at $\varepsilon=0$.  The absolute value of the
difference of the adsorption strength at these two intersection points (i.e.,
difference of their abscissa values) defines the transition width
\begin{equation}
\delta\varepsilon=\Big|\frac{Z_{\mathrm{end}}|_{\varepsilon=0}}{l}\Big|
\end{equation}

In order to study the transition mechanism we analyze the distribution of
minority chain ends, which we denote as $P_z$. The product $P_z \:
{\rm d}z$ represents the probability to find the free end in a layer with
its $z$ coordinate ranging from $z$ to $z+ {\rm d}z$. In the MC
scheme, $P_z$ is evaluated as follows: We split the simulation box
along the $z$ direction into $n_z$ layers, where each layer has the volume
$L_x*L_y*L_z/n_z$ and an index $k$ ranging from $0$ to $n_z-1$.  Then $P_z(k)$
is obtained by counting how often the free end was located in the $k$th layer,
divided by ${\rm d} z$ and by the total size of the sample.

\begin{figure}[h]
  \centering
  \subfigure[]{
    \label{Fig:Pz} 
    \includegraphics[angle=0, width=7.0cm]{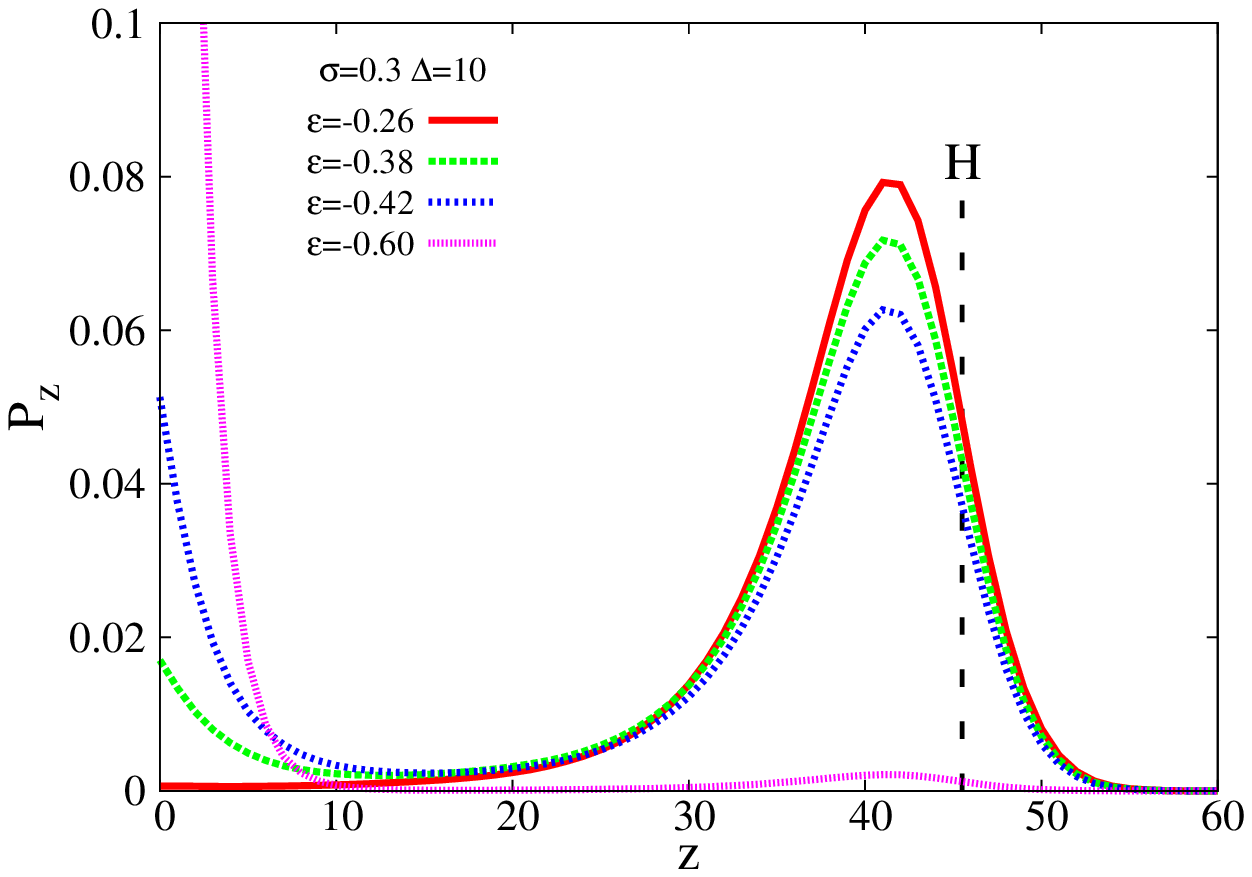}}
  \subfigure[]{
    \label{Fig:freeEnergy} 
    \includegraphics[angle=0, width=7.0cm]{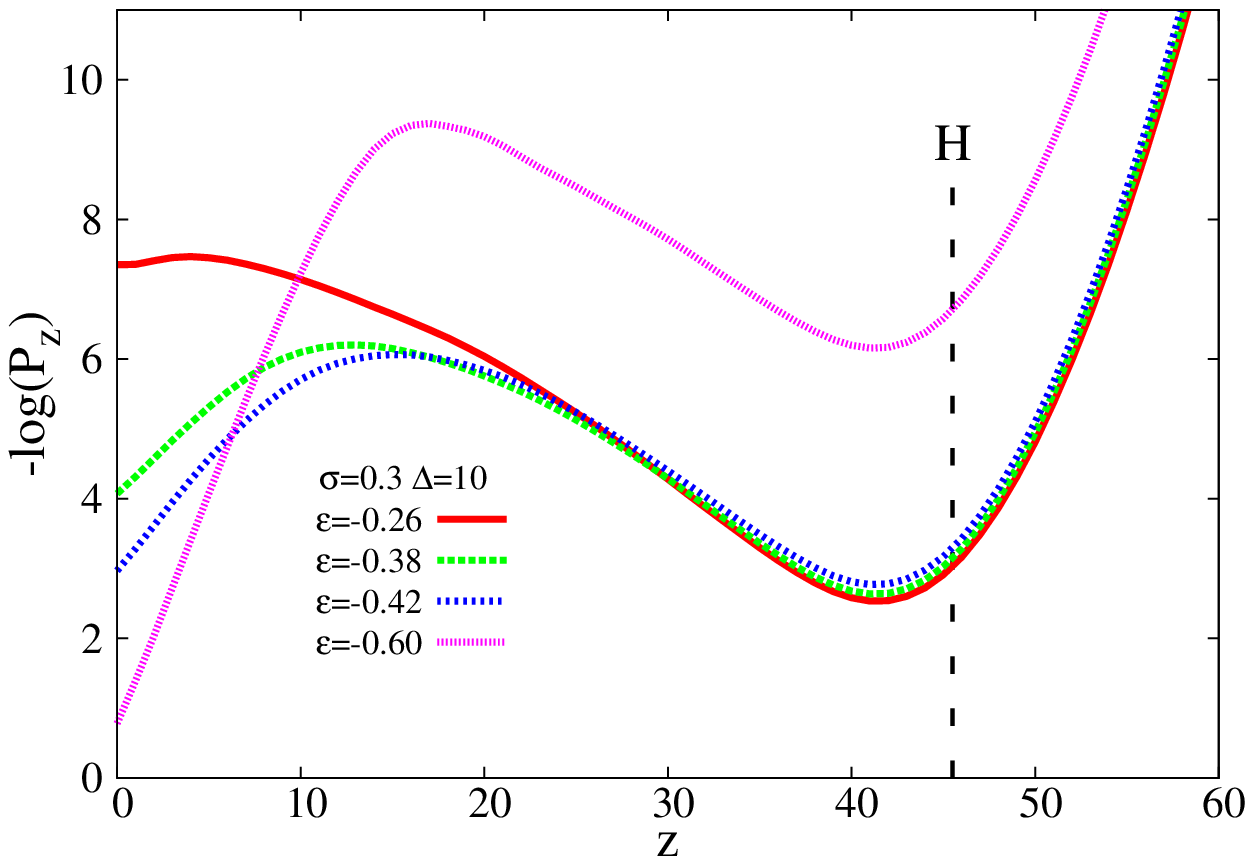}}
  \caption{
Longitudinal distribution of the free end of the minority chain obtained 
from MC simulations as functions of $z$ at $\sigma=0.3$, $\Delta=10$, for different
adsorption strength (a). The Landau free energy is obtained by taking the 
negative logarithm of the distribution $P_z$ in Fig.~\ref{Fig:Pz}. The black dashed
vertical line is shown to denote the brush height $H\approx46$ at $N_b=100$, $\sigma=0.3$.
}
  \label{Fig:end_z} 
\end{figure}

Fig.\ \ref{Fig:Pz} shows examples of distributions $P_z(z)$ for a minority
chain of length $N=110$ in a brush with grafting density $\sigma=0.3$. One can
clearly see that two populations of different conformations (adsorbed and
exposed) are involved in the transition. In the transition region these
populations coexist which is reflected by a bimodal distribution. Specifically
one can see that at low adsorption strength (e.g., $\varepsilon=-0.26$), the
chain end is mostly located outside of the brush, whereas at high adsorption
strength (e.g., $\varepsilon=-0.6$), the chain end is localized close to the
surface. At intermediate adsorption strengths ($\varepsilon\sim-0.42$), the
adsorbed and the exposed state coexist and the minority chain switches back and
forth from one to the other.

A conventional way to analyze a phase transition is to construct a ``Landau free energy"
from the logarithm of the statistical distribution of the order parameter \cite{nonequilibrium}. 
We will use the position $z$ of the free end of the minority chain
as the order parameter.  In this case the Landau free energy is $-\ln P_z(z)$.
Selected free energy curves are shown in Fig.\ \ref{Fig:freeEnergy}. The free
energy as a function of chain end position has one minimum close to $z=0$ for
large $\varepsilon$ (adsorbed state), and one minimum close to $z=40$ for small
$\varepsilon$ (exposed state). For intermediate $\varepsilon$, these two minima
compete with each other. An alternative definition of the transition point
$\varepsilon^{*}$ is the point where both minima have equal depth. However, we
wish to stress that at finite chain length, there is no classical phase
transition, but rather a crossover point between two regimes.

The height of the free energy barrier between the exposed state and the
adsorbed state, $U_{\mathrm{barrier}}$, is extracted from the Landau function
curve at this transition point.  The equal depth condition for the transition
point makes the barrier height definition unambiguous. In the following we
compare the transition characteristics as obtained by simulations to the
scaling prediction of the analytical theory laid out in Section
\ref{sec:theory}.

\begin{figure}[h]
  \centering
   {\includegraphics[angle=0, width=7.0cm]{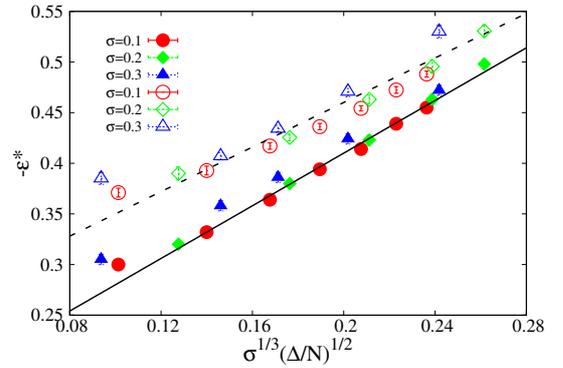}}
  \caption{Transition points as a function of the scaling parameter
$\sigma^{1/2}(\Delta/N)^{1/2}$ obtained from finding the value $\varepsilon^*$
corresponding to the criterion of equal height in $P_z$ (see Fig.\
\ref{Fig:Pz}) (filled symbols), or to the criterion of maximum slope of
$Z_\mathrm{end}$ (see Fig.\ \ref{Fig:Zend_MC}) (empty symbols). The solid and
dash lines are used to guide the eyes and they have almost the same slope}
  \label{Fig:MC_transition_point} 
\end{figure}

Fig.\ \ref{Fig:MC_transition_point} displays the transition point, i.e. the
adsorption energy $-\varepsilon^{*}$ as a function of the scaling parameter
$\sigma^{1/3}\left(\frac{\Delta}{N}\right)^{1/2}$ combining the grafting
density, and the ratio of the minority-to-majority chain lengths. Two
definitions of $-\varepsilon^{*}$ have been proposed above, and data points
corresponding to both definitions are presented.  Each set collapses into a
single master curve close to the straight line suggested by Eq.\
(\ref{eq:transition_point}) but due to the effect of finite chain length
the lines are separate. We expect the difference to vanish in the appropriate
thermodynamic limit $N\to\infty, \:N_{b}\to\infty, \:\frac{\Delta}{N}=const,
\:\sigma=const$, where the transition becomes truly sharp. The general trend is
that the transition point shifts to larger adsorption strength $|\varepsilon^*|$
with increasing  minority chain length and grafting density. This is to be
expected, since both factors stabilize the exposed state against the
competing adsorbed state.

\begin{figure}[h]
  \centering
  \subfigure[]{
    \label{Fig:sharpness_raw} 
    \includegraphics[angle=0, width=7.cm]{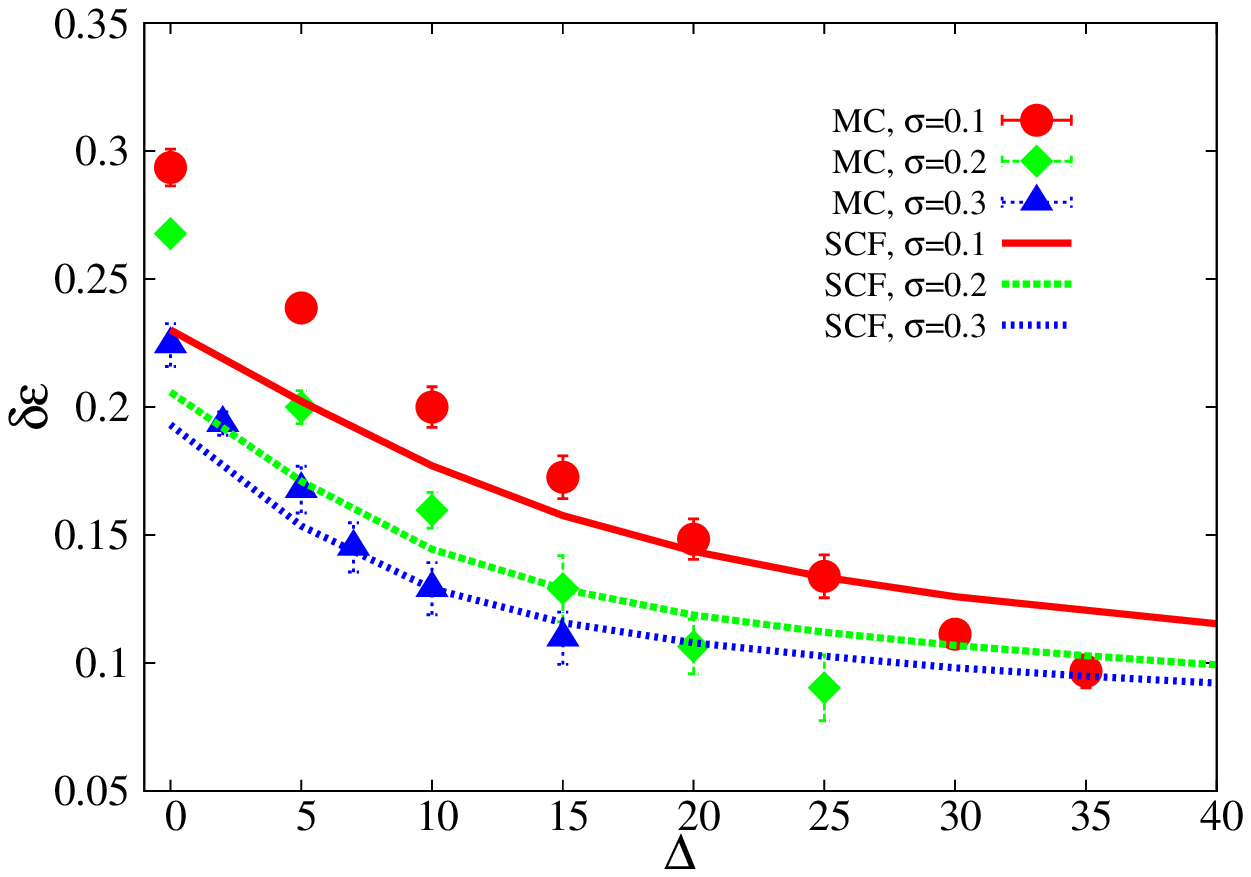}}
  \subfigure[]{
    \label{Fig:barrier_raw} 
    \includegraphics[angle=0, width=7.cm]{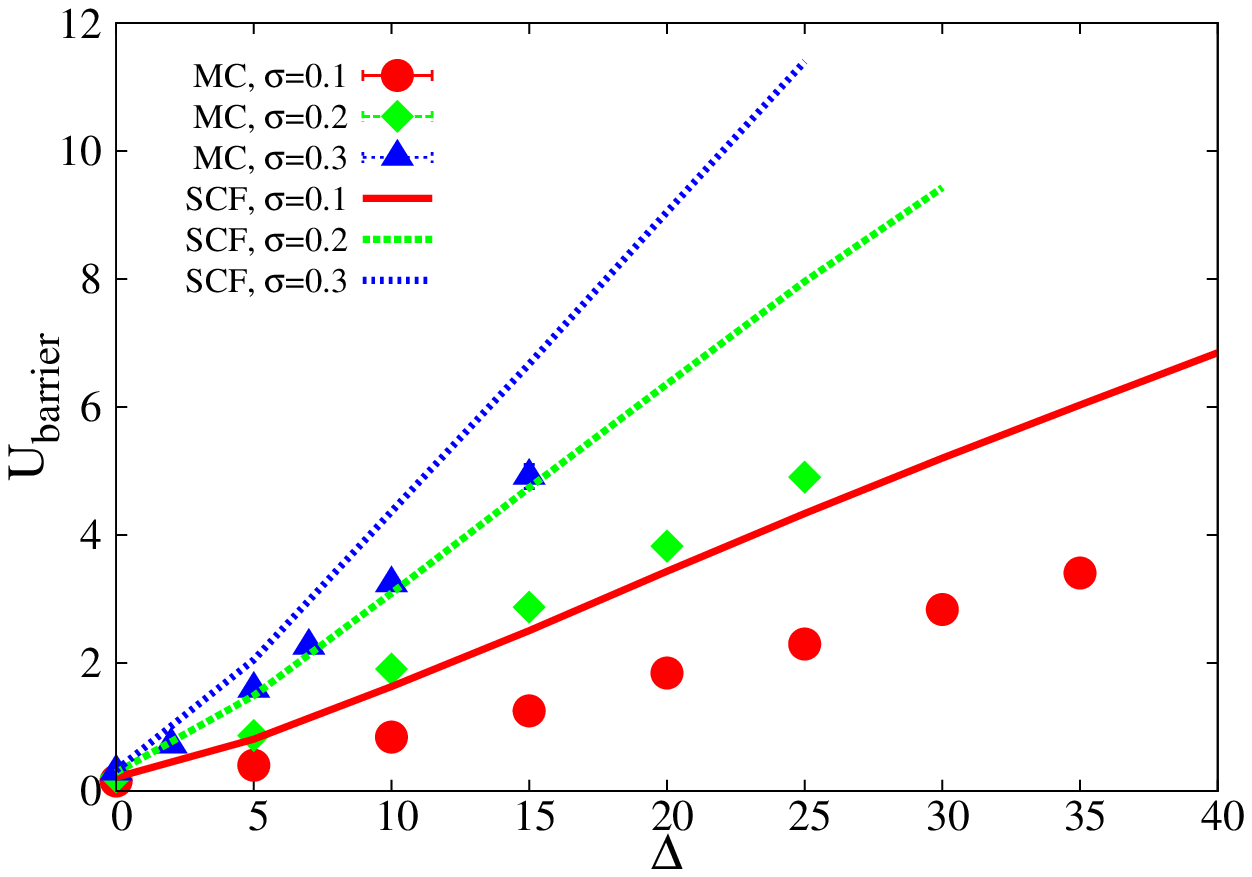}}
  \caption{Transition sharpness (a) and barrier height between coexisting states (b)
as a function of the chain length 
difference $\Delta =N-N_b$ between minority chain and brush
chains for different grafting densities as indicated, according
to SCF theory (lines) and MC simulations (symbols).
}
  \label{Fig:transition_rawdata} 
\end{figure}

In Fig.\ \ref{Fig:MC_transition_point}, the intercept and the slope are linked
to the model-specific dependence of the adsorbed chain's free energy on
$\varepsilon$, which were therefore treated as free parameters by the theory.
One has to be cautious, however, not to overestimate the agreement between
theory and simulations. The scaling predictions rely on two assumptions: (i)
logarithmic corrections are neglected, implying that the relevant free energies
per chain are much larger than 1 (in $kT$ units), $V_{0}\Delta\gg1$; (ii) the
adsorption is assumed to be weak, such that the expansion of $\mu(\varepsilon)$
in Eq.\ (\ref{eq:mu_epsilon}) can be applied, $\frac{V_{0}\Delta}{N}\ll1$.
At brush chain lengths $N \approx N_{b}=100$, it is difficult to satisfy
both conditions simultaneously. In systems with small values of $\Delta$
and/or small $\sigma$, the first condition becomes questionable, whereas in
systems with larger $\Delta$ and/or larger $\sigma$, the second condition may
be violated. Therefore, the observed reasonably good collapse onto a straight
line might be the result of cancellation effects.

The two most important characteristics of the transition, as far as potential
applications are concerned, are the transition sharpness (or effective width)
and the waiting time characterizing the transition kinetics. The present work
does not include dynamic simulations, so we use the free energy barrier at the
transition as an indirect measure of the waiting time. The transition width,
denoted as $\delta\varepsilon$ and characterized by the inverse slope of
$Z_{\mathrm{end}}(\varepsilon)$, is plotted in Fig.\ \ref{Fig:sharpness_raw} as
a function of the minority chain length increment $\Delta$, while the barrier
height (in units of $kT$) is displayed in the Fig.\ \ref{Fig:barrier_raw}. We
present both the MC results (symbols) and the SCF results (lines). The
transition becomes sharper for longer minority chains and larger grafting
densities. The difference between the MC and the SCF results for $\delta
\varepsilon$ (Fig.\ \ref{Fig:sharpness_raw}) is noticeable but not dramatic.
As discussed in Sec.\ \ref{sec:theory} (Eq.\ (\ref{eq:sharpness scaling})),
the transition sharpness is essentially determined by the response of
the free energy of the adsorbed state to variations of the adsorption
strength $\varepsilon$, which does not seem to be affected by fluctuations
very much. In contrast, the effect of fluctuations on the free energy
barrier is quite strong.
The free energy barrier grows approximately linearly with $\Delta$ and also
increases with the grafting density. The absolute values of the
barrier height obtained by MC simulations are smaller than those from SCF
method by roughly a factor of 2 (Fig.\ \ref{Fig:barrier_raw}). This is clearly
a pronounced effect of density fluctuations implying strong consequences on the
transition kinetics.

\begin{figure}[h]
  \centering
  \subfigure[]{
    \label{Fig:sharpness_scaled1} 
    \includegraphics[angle=0, width=7cm]{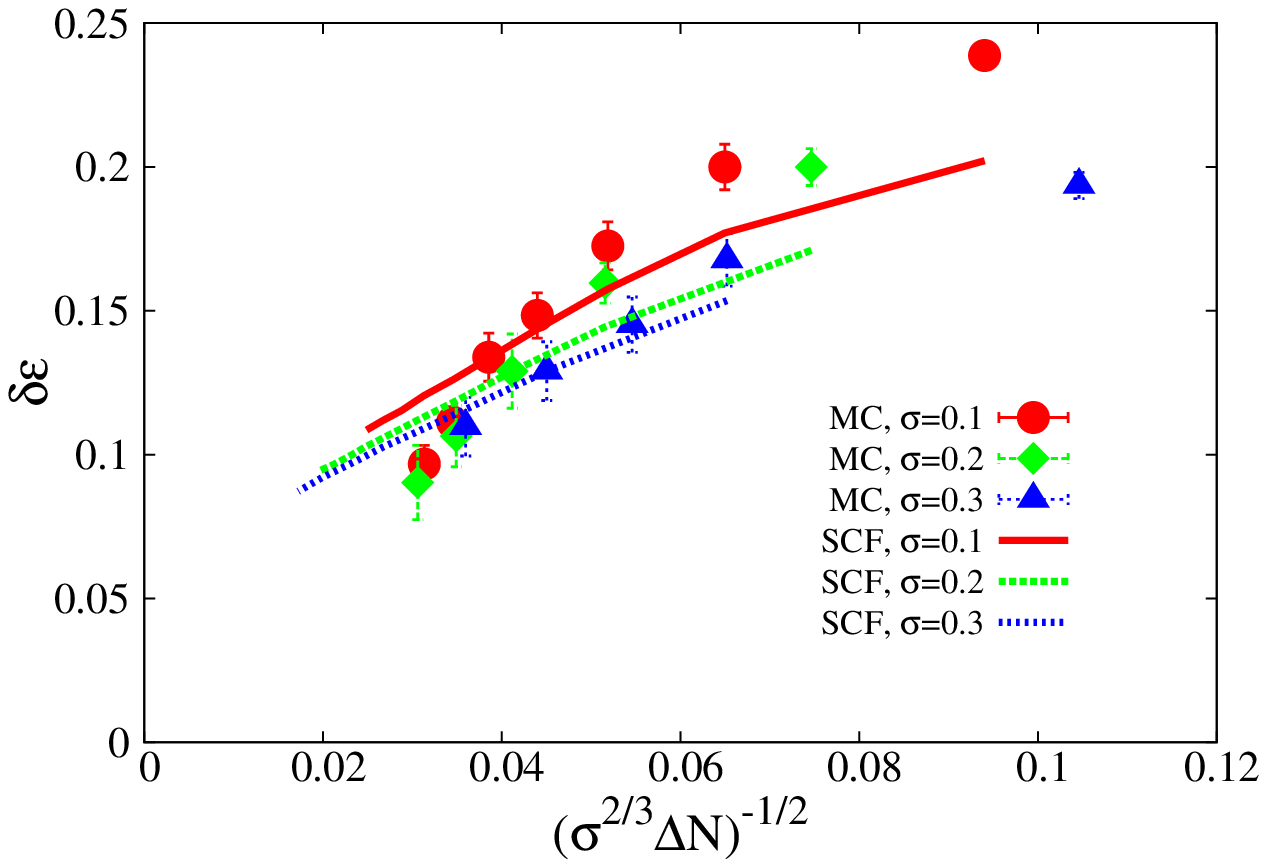}}
  \subfigure[]{
    \label{Fig:barrier_scaled1} 
    \includegraphics[angle=0, width=7cm]{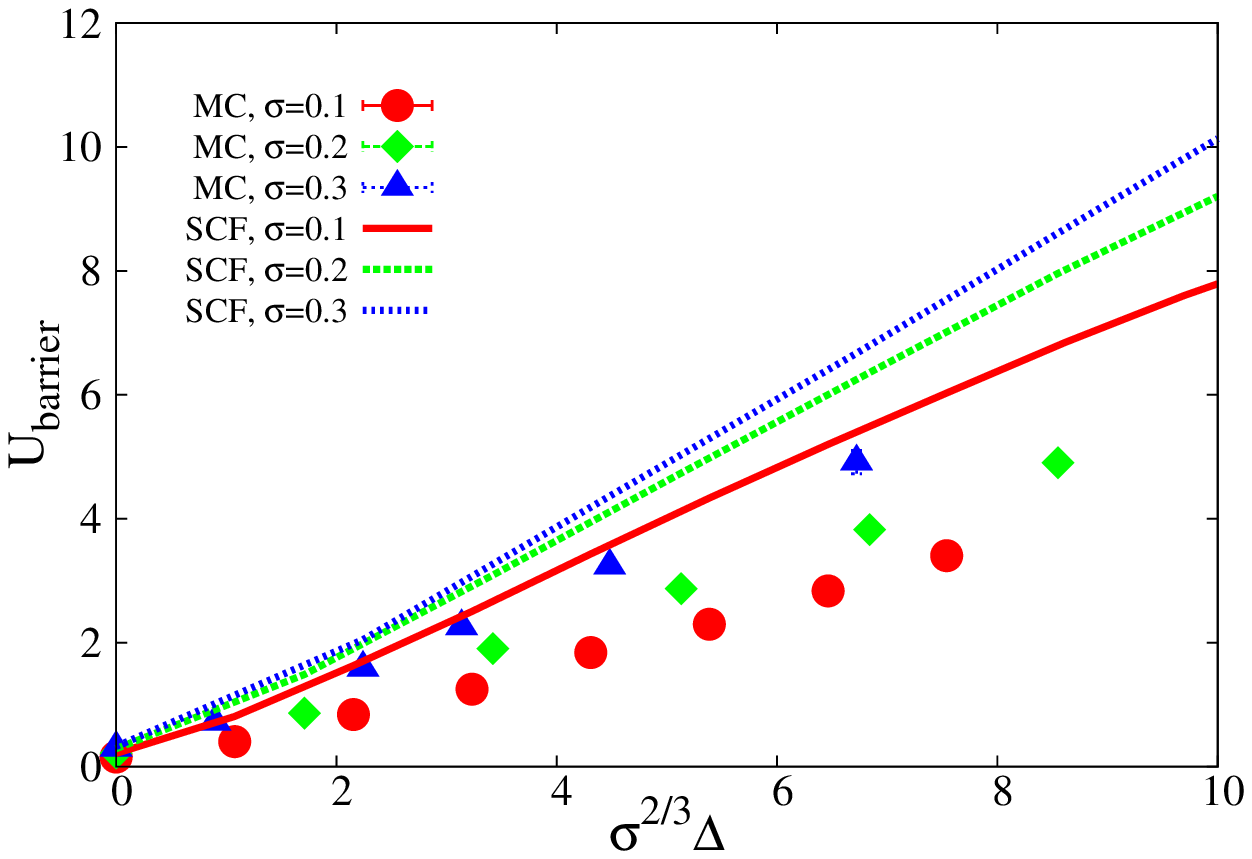}}
  \caption{
Same data as Fig.\ \protect\ref{Fig:transition_rawdata},
shown in the scaling plot suggested by the analytical 
predictions in Sec.~\ref{sec:theory}
with the scaling variable $(\sigma^{2/3} \Delta)$.
}
  \label{Fig:transition_scaled1} 
\end{figure}

The same data are replotted in Fig.\ (\ref{Fig:transition_scaled1}) with
scaling coordinates as suggested by the analytical theory, see Eqs
(\ref{eq:sharpness scaling}),(\ref{eq:barrier scaling}).  The curve for
$\delta\varepsilon$ vs. $\sigma^{-1/3}\left(N\Delta\right)^{-1/2}$ is expected
to pass through the origin since in the thermodynamic limit, one has
$\sigma^{-1/3}\left(N\Delta\right)^{-1/2}\rightarrow0$, and the transition
becomes jump-like with $\delta\varepsilon\rightarrow0$.  At finite brush
chain length $N_b$ and for larger values of the scaling parameter
$\sigma^{-1/3}\left(N\Delta\right)^{-1/2}$ finite chain length effects and
logarithmic corrections become important.  This presumably explains why the
data do not collapse on a master curve, and why the SCF data don't even
seem to converge towards the expected behavior $\delta \varepsilon \rightarrow
0$ at $\sigma^{-1/3}\left(N\Delta\right)^{-1/2}\rightarrow0$.  The MC data
are better compatible with a limiting straight line passing through the origin,
and the slope has the correct order of magnitude.  According to the theoretical
expectation (Eqs (\ref{eq:transition_point}) and (\ref{eq:sharpness scaling}))
the slope in Fig.\ \ref{Fig:sharpness_scaled1} should exceed that in
Fig.\ \ref{Fig:MC_transition_point} by a factor of $\frac{2}{\kappa}\simeq1.2$.
The actual slopes differ by a factor of 2.3. 

Fig.\ \ref{Fig:barrier_scaled1} displays the energy barriers at the
transition as a function of $\sigma^{2/3}\Delta$. The SCF data are quite close
to collapsing onto a single straight line, as predicted by Eq.\
(\ref{eq:barrier scaling})  while the MC data are not. Although the direct
proportionality to the chain length increment $\Delta$ is still satisfied
reasonably well, the effect of increasing the grafting density is stronger than
predicted by the theory. This is probably due to the fact that the values of
the barrier height $U_{\mathrm{barrier}}$ obtained from MC simulations lie in
the range between 0.5 and 5 $k_B T$, whereas the theory assumes that the two
states are well separated, implying $U_{\mathrm{barrier}}/k_B T \gg 1$.  Thus
the asymptotic behavior assumed by the theory is not yet reached, which may
explain the poor data collapse.

\begin{figure}[h]
  \centering
   {\includegraphics[angle=0, width=7.0cm]{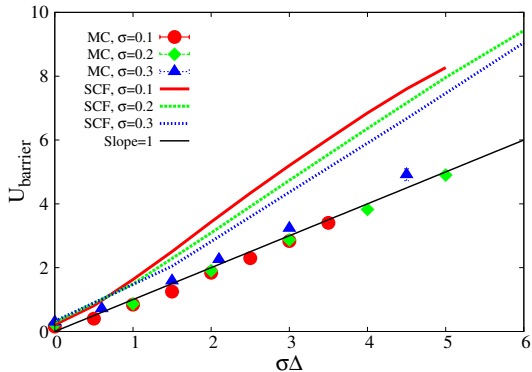}}
  \caption{
Same data of the transition barrier as Fig.\
\protect\ref{Fig:transition_scaled1}, scaled as suggested by a scaling
hypothesis with modified scaling variable $\sigma\Delta$.  
}
\label{Fig:barrier_scaled2} 
\end{figure}

It is worth noting that the data do seem to obey an apparent scaling
$U_{\mathrm{barrier}}\sim\sigma\Delta$ as illustrated by Fig.\
\ref{Fig:barrier_scaled2}. 
We are not aware of a theoretical reason for such a scaling. 
The observed apparent scaling might be the purely fortuitous result
of various corrections to the analytical theory, which relies on
several somewhat questionable assumptions as discussed above.

\section{Conclusion and remarks}\label{sec:conclusions}

To summarize, we have investigated in detail conformational properties and
phase transition behavior of a type of stimuli-responsive polymer materials
using particle-based Monte Carlo simulations and continuum-based 3-dimensional
self-consistent field theory. This type of stimuli-responsive materials is
designed based on a polymer brush with a small fraction of adsorption-active
minority chains introduced in the brush, and the minority chain can work as a
responsive sensor. The basic mechanism for this responsive sensor relies on the
conformational switch of the minority chain. The adsorption between the
substrate and the minority chain serves as the trigger that switches the chain
conformations. In practice, the trigger could also be a temperature or solvent
composition change. Upon varying the adsorption strength, two states were
observed, one being an exposed state, in which the free end of the minority
chain is located at the outside surface of the brush and possesses a
stem-crown-like configuration, and the other being an adsorbed state, in which
the minority chain is located at the adsorbing substrate in a nearly
2-dimensional confinement.

\begin{figure}[h]
\centerline{\includegraphics[angle=0,scale=0.6,draft=false]{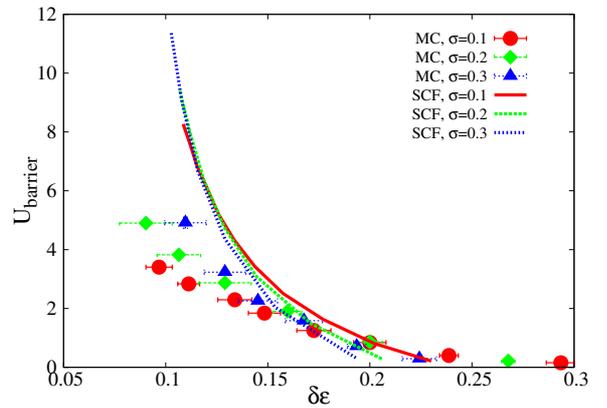}}
\caption{
Barrier height versus sharpness of the transition according
to SCF calculations (lines) and MC simulations (symbols).
At given sharpness $\delta \varepsilon$, the barrier height
is significantly smaller in the MC simulations.
}
\label{Fig:barrier_vs_sharpness} 
\end{figure}

One important purpose of the present study was to highlight the effect of
fluctuations by comparing two numerical methods applied to one and the same
model, namely MC simulations and 3d SCF calculations. The comparison shows that the SCF theory roughly captures the sharpness of
the transition (Fig.\ \ref{Fig:sharpness_scaled1}, but overestimates the
barrier height (Fig.\ \ref{Fig:barrier_scaled1}). As discussed earlier, density
fluctuations at the outer brush surface (see Fig.\ \ref{Fig:density_scaled})
effectively reduce the range of the repulsive barrier created by the brush and
hence the degree of chain stretching in the exposed state (see Fig.\
\ref{Fig:Zend_MC_SCF}). As a consequence, the free energy barrier between the
adsorbed and the exposed state is also reduced.  The practical importance of
this effect can be seen when plotting the barrier height against the sharpness
of the transition (Fig.\ \ref{Fig:barrier_vs_sharpness}).  The transition
sharpness is a measure for the sensitivity of the switch to small changes in
the environment. It can be tuned by adjusting the minority chain length or the
grafting density.  Assuming that it has been tuned to a certain value, 
Fig.\ \ref{Fig:barrier_vs_sharpness} demonstrates that the corresponding
barrier height between coexisting states at the transition is greatly reduced
in the MC simulations compared to the SCF prediction.  Since the barrier height
determines the time required for switching between states, this can drastically
reduce the response times of the switch.  For example, for $\delta \varepsilon =
0.12$, the reduction can be as high as $7 k_B T$, and according to a simple
Arrhenius-type estimate this would lead to a speedup of the switching time by
three orders of magnitude.

The switching time is one major issue for the quality of sensors or other
responsive materials. In our previous work \cite{Klushin_switch}, we
performed overdamped Brownian dynamics simulations of a single switch chain in
a static brush potential to estimate the switching time.
By tracking the position of the free end bead of the
minority chain, we found that this bead could jump between two well separated
positions (which correspond to the exposed and adsorbed state, respectively)
during a time about $10^6\tau_\mathrm{mono}$, where 
$\tau_\mathrm{mono}$ is the characteristic relaxation time of a single
monomer, which is typically on the order of $10^{-9}$ seconds for flexible 
chains. We thus estimated the switching time to be on the
order of milliseconds. However, in that calculation, the brush was represented
in a simplified manner by a parabolic profile and the coupling between the
minority chain and the brush chains was neglected, which presumably leads to a
large deviation from the true switch time. Molecular dynamic simulations that
model a brush-minority chain system in a dynamic manner are clearly desirable.

In the present study, we restrict ourselves to systems with good solvent
conditions, and assume that two-body interactions dominate, thus we only
consider excluded volume interactions.
However, for dense brushes, higher order contributions will become
important. Higher order terms in the free energy virial expansion provide
corrections that lead to some change in the brush height and in the overall
free energy of the competing states. Hence one would expect corrections to
the properties of the switching transition, in particular the barrier height at
transition. In the case of very dense grafting these corrections may be
significant but this falls outside the scope of the present paper.

Besides fluctuations, another factor that must be taken into account in real
systems is polydispersity. In the present study, the brush polymers were
taken to be monodisperse. Real polymer brushes, however, are always
polydisperse, and it has been demonstrated that polydispersity significantly
alters both the equilibrium properties \cite{poly1, poly2} and dynamical
properties \cite{poly_dynamics1,poly_dynamics2,poly_dynamics3} of materials.
Using a one-dimensional SCF theory, we have performed a preliminary study of a
system with an adsorption-active chain embedded in a polydisperse brush with a
continuous Schulz-Zimm length distribution. We found that for $\Delta\gtrsim
15$, the switch sharpness is similar to that of the monodisperse brush obtained
by 1-dimensional SCF theory; however, the switch barrier is reduced. This
suggests that the switching mechanism should be at least as effective in
polydisperse brushes than in monodisperse brushes, the switching might even be
faster.  Further and more detailed studies are currently under way.

An important question pertaining to the MC scheme used in our work is the
choice of the size of the density averaging cell, $b$, which has the meaning of
an interaction range. In section \ref{sec:brush} we have shown that the
effective excluded volume parameter appears as the result of a renormalization
of the ``bare'' parameter due to density fluctuations. For a given set of
parameters $N_{b}$ and $\sigma a^{2}$, where $a$ is the statistical segment
length unit, the brush profile is uniquely defined by $v_\mathrm{eff}$ or,
more precisely, the corresponding second virial coefficient in the MC model,
$B_\mathrm{eff}$, while the renormalization factor 
$\frac{B_\mathrm{eff}}{B_\mathrm{bare}}$ was shown
to be a function of a single dimensionless variable $\frac{b}{a}$. The
renormalization of monomer interaction parameters has been studied for single
flexible swollen chains \cite{grosberg_80}, in the context of dense polymer
systems \cite{Particle_Mesh}, and in the field-theoretical context
\cite{wang_02,kudlay_03,katz_05,morse_07}.  We would like to emphasize that
polymer brushes simulated by MC in the Laradji version \cite{MC_CG} represent a
class of model systems where renormalization effects are quite strong and can
be extracted easily. The basic tenet is that large-scale properties should
depend only on the renormalized excluded volume parameter while its bare value
may become relevant on small length scales comparable to $a$ or $b$.  One
consequence in our system is that the brush profiles very close to the outer
surface cannot be described by a single renormalized $v_\mathrm{eff}$ (see
Fig.\ \ref{Fig:density_scaled}), and that the surface interaction parameter
$\varepsilon$ has to be renormalized independently from $v_\mathrm{bare}$ (as has been done
in Fig.\ \ref{Fig:Zend_MC_SCF}).

However, even after taking into account the renormalization of interaction
parameters, we find that the properties of chains in close vicinity and above
the brush surface are still strongly influenced by density fluctuations. 
The density fluctuations at the brush surface significantly reduce
the free energy barrier between the exposed and the adsorbed state in Fig.\
\ref{Fig:barrier_vs_sharpness}. Thus, the minority chain serves as a delicate
probe of both local and large-scale fluctuation effects.

\bigskip
\begin{center}
\textbf{ACKNOWLEDGMENTS}
\end{center}

Financial support by the Deutsche Forschungsgemeinschaft (Grants No. SCHM
985/13-1, and No. 13-03-91331-NNIO-a) is gratefully acknowledged. S. Q
acknowledges support from the German Science Foundation (DFG) within project C1
in SFB TRR 146. Simulations have been carried out on the computer cluster Mogon
at JGU Mainz.

\appendix
\section{SCF theory}

The SCF method used in the present work is formulated in the continuous 3d
space, starting from the same Hamiltonian with that in the MC method mentioned
in the text. In the SCF approach \cite{SCF_R1,SCF_R2,SCF_book}, the partition
function $Z=\sum_{\{\mb R\}}\exp(-\beta\mc H(\{\mb R\}))$ is rewritten as a
functional integral, i.e., $Z=\int\mc D\omega\mc D\rho_t\exp(-\beta\mc F)$,
with the free energy functional expressed as
\begin{eqnarray}
\beta\mc F&=&\frac{v}{2}\int d\mb r\rho^2_t(\mb r)
 -\int d\mb r\omega(\mb r)\rho_t(\mb r)\nonumber\\
&-&n_b\ln Q_b[\omega]-\ln Q_m[\omega+U_\mathrm{ads}],
\end{eqnarray}
where $\rho_t$ is the total density distribution, $\omega$ is the corresponding
conjugate auxiliary potential, and $Q_b$, $Q_m$ are the single chain partition
functions. In the calculation we set the excluded volume parameter 
$v=v_\mathrm{eff}$ (where $v_\mathrm{eff}$ is obtained from the MC simulation).
Extremizing the free energy functional with respect to the
fluctuating fields, i.e., the density $\rho_t$ and the potential $\omega$,
one obtains the SCF equations
\begin{eqnarray}
\omega&=&v\rho_t\nonumber\\
\rho_t&=&\rho_b+\rho_m =
 \int_0^{N_b}dsq_b^\dagger(\mb r,s)q_b(\mb r,N_b-s)\nonumber\\
&+&
\int_0^Ndsq_m^\dagger(\mb r,s)q_m(\mb r,N-s)
\end{eqnarray}
where $q_\mu^\dagger$ and $q_\mu$ ($\mu=b,m$) satisfy the modified diffusion equation
\begin{eqnarray}
\frac{\partial}{\partial s}q_\mu(\mb r,s)
  &=&\frac{a^2}{6}\nabla^2q_\mu(\mb r,s)-\omega_\mu q(\mb r,s), 
\\
\frac{\partial}{\partial s}q_\mu^\dagger(\mb r,s)
  &=&\frac{a^2}{6}\nabla^2q_\mu^\dagger(\mb r,s)-\omega_\mu q(\mb r,s), 
\end{eqnarray}
with $\omega_\mu=\omega$ for $\mu=b$, $\omega_\mu=\omega+U_\mathrm{ads}$ 
for $\mu=m$, Dirichlet boundary conditions along the $z$ direction, and
periodic boundary conditions along the $x$ and $y$ directions.
The functions $q_\mu$ and $q_\mu^\dagger$ ($\mu=b,m$) correspond
to spatial integrals of the single-chain propagator between monomer $s$ and 
the free or grafted end, respectively, over all possible end positions.
Thus the initial condition for $q_\mu$ (free end) is $q_\mu(\mb r,0)=1$,  
and the initial condition for $q_\mu^\dagger$ (grafted end) is chosen
self-consistently such that the density of graft points is reproduced 
correctly, i.e.,
$q_b^\dagger(\mb r,0) \: q_b(\mb r, N_b) = \sigma \delta(z-z_0) $
for the brush chains, and
$q_m^\dagger(\mb r,0) \: q_m(\mb r, N) = \delta(\mb r - \mb r_0) $
for the minority chain, where $\mb r_0 = (0,0,z_0)$ is positioned
at the center of the substrate. The transition properties are extracted from
the corresponding minority chain propagator. For example, one can calculate
the distribution of free end via $P_\mathrm{end}(z)=\int {\rm d}x \: {\rm d} yq^\dagger_m(\mb r, N)/\int d\mb
rq_m^\dagger(\mb r,N)$. From these distributions the transition point and transition barrier can be obtained in a similar way
as that used in the MC simulations.

The SCF equations are closed and iterative methods are used to find their
solutions. The auxiliary potential $\omega$ is used as the iterative variable,
and a new potential for the next iteration is generated
$\omega^{(n+1)}=\omega^{(n)}+\lambda(v\rho_t^{(n)}-\omega^{(n)})$, where
$\lambda$ is a small number controlling the stability of the iteration and the
speed of converging. In most of the cases we use $\lambda=0.03$. The iteration
process stops if the iteration error is smaller than $10^{-5}$. There are
several ways to solve the modified diffusion equation numerically, for example,
real space finite difference schemes \cite{CN1}, pseudo-spectral schemes
\cite{pseudo1, pseudo2}, and so forth. Usually, pseudo-spectral schemes are
reliable and efficient. However, in our system, we encountered numerical
problems when chains are strongly stretched (e.g. at large grafting densities,
or when chains are long). In such cases, the propagators sometimes assumed
negative values, and the SCF equations did not converge.  Similar problems were
reported in Refs.\ \cite{SCF4,SCF5}.  Therefore, we adopt the Doublas-Brian
scheme \cite{book_DB}, a real-space finite-difference method which did not
suffer from this problem.  The volume of the simulation system is chosen equal
to that of the MC method. It is divided uniformly into $n_x*n_y*n_z=40*40*200$
grid cells, and continuous quantities are discretized on the vortices of these
cells. The chain contour including the minority chain is uniformly discretized
into 400 steps.  About 300 hundred iteration steps are necessary for the
iteration to reach convergence.

\end{document}